\documentclass[useAMS,usenatbib]{mn2e}
\usepackage{psfrag,graphicx}
\usepackage{color}
\usepackage{txfonts}
\usepackage{breqn}
\usepackage[T1]{fontenc}
\usepackage{ae,aecompl}

\title[Black hole formation in the early universe]
{Assessing inflow rates in atomic cooling halos: implications for direct collapse black holes}
 \author[Latif \& Volonteri]
{M. A. Latif\thanks{Corresponding author: latif@iap.fr}$^{1,2}$,
M.  Volonteri$^{1,2}$ \\
$^1$ Sorbonne Universités, UPMC Univ Paris 06, UMR 7095, Institut d'Astrophysique de Paris, F-75014, Paris, France \\
$^2$CNRS, UMR 7095, Institut d'Astrophysique de Paris, F-75014, Paris, France
  }
\date{}

\def\LaTeX{L\kern-.36em\raise.3ex\hbox{a}\kern-.15em
      T\kern-.1667em\lower.7ex\hbox{E}\kern-.125emX}

\begin{document}

\bibliographystyle{mn2e}

\label{firstpage}

\maketitle
\def\na{NewA}%
\def\aj{AJ}%
\def\araa{ARA\&A}%
\def\apj{ApJ}%
\def\apjl{ApJ}%
\def\jcap{JCAP}

\def\apjs{ApJS}%
\def\ao{Appl.~Opt.}%
\def\apss{Ap\&SS}%
\def\aap{A\&A}%
\def\aapr{A\&A~Rev.}%
\def\aaps{A\&AS}%
\def\azh{AZh}%
\def\baas{BAAS}%
\def\jrasc{JRASC}%
\def\memras{MmRAS}%
\def\mnras{MNRAS}%
\def\pra{Phys.~Rev.~A}%
\def\prb{Phys.~Rev.~B}%
\def\prc{Phys.~Rev.~C}%
\def\prd{Phys.~Rev.~D}%
\def\pre{Phys.~Rev.~E}%
\def\prl{Phys.~Rev.~Lett.}%
\def\pasp{PASP}%
\def\pasj{PASJ}%
\def\qjras{QJRAS}%
\def\skytel{S\&T}%
\def\solphys{Sol.~Phys.}%

\def\sovast{Soviet~Ast.}%
\def\ssr{Space~Sci.~Rev.}%
\def\zap{ZAp}%
\def\nat{Nature}%
\def\iaucirc{IAU~Circ.}%
\def\aplett{Astrophys.~Lett.}%
\def\apspr{Astrophys.~Space~Phys.~Res.}%
\def\bain{Bull.~Astron.~Inst.~Netherlands}%
\def\fcp{Fund.~Cosmic~Phys.}%
\def\gca{Geochim.~Cosmochim.~Acta}%
\def\grl{Geophys.~Res.~Lett.}%
\def\jcp{J.~Chem.~Phys.}%
\def\jgr{J.~Geophys.~Res.}%
\def\jqsrt{J.~Quant.~Spec.~Radiat.~Transf.}%
\def\memsai{Mem.~Soc.~Astron.~Italiana}%
\def\nphysa{Nucl.~Phys.~A}%
\def\physrep{Phys.~Rep.}%
\def\physscr{Phys.~Scr}%
\def\planss{Planet.~Space~Sci.}%
\def\procspie{Proc.~SPIE}%

%


 \begin{abstract}
 {
Supermassive black holes  are not only common in the present-day galaxies, but billion solar masses black holes also powered $z\geq 6$ quasars.  One efficient way to form such black holes is  the  collapse of a massive  primordial gas cloud into a so-called direct collapse black hole. The main requirement for this scenario is the presence of large accretion rates of $\rm \geq 0.1~M_{\odot}/yr$ to form a supermassive star. It is not yet clear how and under what conditions such accretion rates can be obtained. The prime aim of this work is to determine the mass accretion rates under non-isothermal collapse conditions. We perform high resolution cosmological simulations for three  primordial halos of a few times $\rm 10^7~M_{\odot}$ illuminated by an external UV flux, $\rm J_{21}=100-1000$. We find that  a rotationally supported structure of about parsec size is assembled, with an aspect ratio between $\rm 0.25 - 1$ depending upon the thermodynamical properties. Rotational support, however, does not halt collapse, and mass inflow rates of $\rm \sim 0.1~M_{\odot}/yr$ can be obtained in the presence of even a moderate UV background flux of strength $\rm J_{21} \geq 100$. To assess whether such large accretion rates can be maintained over longer time scales, we employed sink particles, confirming the persistence of accretion rates of $\rm \sim 0.1~M_{\odot}/yr$.  We propose that complete isothermal collapse and molecular hydrogen suppression may not always be necessary to form supermassive stars, precursors of black hole seeds.  Sufficiently high inflow rates can be obtained for UV flux $\rm J_{21}=500-1000$, at least for some cases. This value brings the estimate of the abundance of  direct collapse black hole seeds closer to that high redshift quasars. 
 } 
 \end{abstract}



\section{Introduction}
The observations of high redshift quasars at $z \geq 6$ reveal the presence of supermassive black holes of about a few billion solar masses in the early universe \citep{Fan2003,Fan2006,Willot2010,MOrtlock2011,Venemans2013,Wu2015}.  From a theoretical perspective, the assembly of such massive objects within the first billion years after the Big Bang is challenging. How and under what conditions they were  formed is a question of prime astrophysical interest. 

Various models for the formation of supermassive black hole seeds have been proposed \citep{Rees1984,Volonteri2010,Volonteri2012,Haiman2012}. One potential pathway is the collapse of a primordial, i.e., metal-free, ``normal" star into a  stellar-mass black hole.  Earlier numerical simulations performed to study the formation of the first generation of stars \citep{Abel2002,Bromm2002,Yoshida2006}  suggested that the masses of the first stars are of the order of a few hundred solar.  However,   recent simulations found that the protostellar disk forming in the minihalo fragments into multiple clumps \citep{Clark11,Greif12,Stacy2012,LatifPopIII13,Latif2015Disk} and leads to the formation of multiple stars.  In  the case of a single star per halo, they may still reach up to a thousand solar masses at z=25 \citep{Hirano2014}, but feedback from the black hole itself shuts its own accretion and limits its growth \citep{Johnson2007,Alvarez2009}.  Perhaps, as suggested by \cite{Madau2014},  they may still grow  under  prolonged episodes of  super-Eddington accretion.

The second possibility could be  the collapse of a dense stellar cluster into a massive black hole due to relativistic instability \citep{Shapiro1987} or  the formation of a black due to the stellar dynamical processes  in the first stellar cluster \citep{Devecchi2009,Rasio2014ApJ,Katz2015} or even the core collapse of a dense cluster of stellar mass black holes leading to the formation of a massive black hole seed \citep{Davies2011ApJ,Lup2014}.  The expected mass of a seed black hole from these scenarios is  a few thousand solar masses.  

A third mechanism could be the collapse of  a protogalactic gas cloud into  a massive central object, a so-called direct collapse black hole \citep{Loeb1994,Bromm03,Begelman2006,Hosokawa12,Schleicher13}. A detailed discussion of these mechanisms is given in reviews \citep{Volonteri2010,Volonteri2012,Haiman2012}, and we summarize here the main features of relevance for this investigation. The particular direct collapse scenario we address here requires the presence of large accretion rates of about $\rm \geq 0.01-0.1 ~M_{\odot}/yr$ \citep{Begelman2010,Hosokawa2013,Schleicher13,Ferrara14}. This is because such accretion rates are the main pre-requisite to rapidly build a  supermassive star of  about $\rm 10^4 - 10^{5}~M_{\odot}$, which would eventually collapse into a black hole retaining most of the star's mass \citep{1999ApJ...526..941B,2002ApJ...572L..39S}. These large accretion rates can either be obtained through dynamical processes or  thermodynamically, by keeping the gas warm (i.e, with a higher Jeans mass) to avoid fragmentation and star formation. \cite{Begelman2006} proposed that self-gravitating gas in  dark matter halos may loose angular momentum via `bars-in-bars' instabilities and efficiently assemble a central massive star. On the other hand, thermodynamical direct collapse requires the suppression of  molecular hydrogen to avoid fragmentation which leads to a monolithic isothermal collapse. In both cases, the halo has to transfer angular momentum efficiently to avoid collapse to be halted by  the angular momentum barrier. For instance, \cite{Choi2014} have found that angular momentum in isothermal conditions can be transported by non-axisymmetric perturbations.

The conditions in primordial halos are ideal for the formation of  massive stars  due to the lack of efficient coolants in the absence of dust and metals.  However,  the presence of  molecular hydrogen  may lead to efficient fragmentation and star formation. UV stellar radiation can quench $\rm H_2$ formation and consequently fragmentation can be avoided. In fact, numerical simulations show that, in principle,  in the absence of molecular hydrogen, a metal-free halo collapses isothermally via atomic line cooling. Consequently,  a supermassive star of about $\rm 10^5~M_{\odot}$ could be assembled via large accretion rates of about $\rm 1~M_{\odot}/yr$ \citep{Latif2013c,Latif2013d,Johnson2013,Regan2014a}. 

The above  scenario requires a high UV flux to dissociate molecular hydrogen, which depends on the radiation spectra of stellar populations. We  define $J_{crit}$ as the value of the UV background flux above which full isothermal collapse occurs because molecular hydrogen formation is suppressed.  $J_{crit}$ depends on the spectrum of the stars contributing to the background. For  hot stars with $\rm T_{rad} =10^5~K$, such as Population III stars (Pop III), $J_{crit}$ is $\geq 10^4$ \citep[in units of $\rm J_{21}=10^{-21} ~erg/cm^2/s/Hz/sr$][]{Omukai2000, Shang2010,Latif2014UV,Latif2015a}. The second generation of stars is, however, favoured as source of UV radiation for this scenario due to  their longer lives and  higher abundance. If such stars have a softer spectrum, with  $\rm T_{rad} =10^4~K$,  the $J_{crit}$ is 400-700, much reduced compared to the Pop III case. Recently, however, the spectra of Pop II stars were computed employing the stellar synthesis code STARBURST, finding that realistic stellar spectra of Pop II stars can be mimicked by adopting radiation spectra with temperature between  $\rm 10^4-10^5$ K \citep{Sugimura14, Agarwal2015}.  \cite{Latif2015a} found that the value of  UV flux necessary to obtain full isothermal collapse for realistic spectra of Pop II stars is  a few times $\rm10^4 $ in units of $\rm J_{21}$ and becomes constant for $\rm T_{rad}$ between  $\rm 2 \times 10^4-10^5$ K. \cite{Regan2014B} considered an  anisotropic monochromatic radiation source and found  that a small amount of $\rm H_2$ forms due to self-shielding even under very strong flux. They further propose that $\rm J_{21} = 1000$ may be sufficient to form a supermassive star. Similarly, high values of $J_{crit}$ are required for complete isothermal collapse when X-ray ionization heating is included  \citep{Latif2015a,Inayoshi2014Xray}. 

Within a scenario that requires full isothermal collapse, the number density of direct collapse black holes strongly depends on the value of $J_{crit}$.  The estimates of \cite{Dijksta2014} suggest that $J_{crit}$= 1000  seems sufficient to produce the observed number of quasars at z $\geq$ 6. However, the black holes density drops by about a few orders of magnitude below their observed abundance for $J_{crit}$$\rm \geq 10^4$. This makes the direct collapse model infeasible by making their sites extremely rare. Hence, it becomes necessary to explore whether high accretion rates can be maintained under non-isothermal conditions. In fact, the work of \cite{Latif2014ApJ} for soft idealised radiation spectra of $\rm 10^4~K$ suggests that it might be possible to form massive stars even in the presence of a moderate amount of $\rm H_2$.  However, it is still unclear  whether for realistic Pop II spectra high accretion rates of $\rm 0.1 ~M_{\odot}/yr$ can be maintained for  fluxes lower than $J_{crit}$ and what value of the UV flux may be sufficient for this purpose.  If so, are these sites abundant enough to reproduce the number number density of quasars at $z>6$?  We address these questions in the present work.

In this study, we explore the formation of supermassive stars under non-isothermal collapse conditions and determine whether the large inflow rates of $\rm 0.1~M_{\odot}/yr$ necessary for their assembly can be obtained. We perform three-dimensional cosmological simulations  to study the collapse of massive primordial halos of a few times $\rm 10^7~M_{\odot}$  forming at z $\geq$10 and vary the strength of the background UV flux emitted by Pop II stars.  In addition to this, we employ a detailed chemical model which includes all the necessary thermodynamical processes and solve them self-consistently with the cosmological simulations.  We derive the dynamical and thermodynamical properties of these halos. The temporal evolution of the mass inflow rates is computed beyond the initial collapse by employing sink particles.  Our results show that one may not necessarily need complete isothermal collapse to form a supermassive star. This  work has important implications  for the observed abundance of quasars at z $>$6 and  provides a potential pathway for the formation of supermassive black holes at high redshift.

This article is structured as follows. In section 2, we describe the simulations setup and numerical methods.  We present our main results in section 3. In  subsections 3.1, 3.2 and 3.3, we show the structural properties of halos and their time evolution, the expected mass inflow rates and the mass estimates of sinks.  In section 4, we summarise our main findings and conclusions.

 \begin{table*}.
\begin{center}
\caption{Properties of the simulated halos.}
\begin{tabular}{ccccccc}
\hline
\hline

Model	& Mass	 & Redshift  & $J_{21}^{\rm crit}$  & spin parameter\\

No & $\rm M_{\odot} $   & $z$  &  $T_{\rm rad}=2 \times 10^4$  K & $\lambda$ \\
\hline                                                          \\
 		
A     & $\rm 5.6 \times 10^{7}$   &10.59  &20000  &0.034\\
B     & $\rm 4.06 \times 10^{7}$  &13.23  &40000  &0.02\\
C     & $\rm 3.25 \times 10^{7}$  &11.13  &50000  &0.03\\
\hline
\end{tabular}
\label{table1}
\end{center}
\end{table*}

\section{Numerical Methods}
The simulations presented in this study adopt the open source three dimensional code enzo version 2.4\footnote{http://enzo-project.org/, changeset:48de94f882d8} \citep{Enzocode2014}.  Enzo is an adaptive mesh refinement (AMR), parallel, grid based, cosmological hydrodynamical simulations code. It utilises the  message passing interface (MPI) library for parallelisation and is well suited for the current simulations.  The hydrodynamical equations are solved using the  finite difference scheme by employing the piece-wise parabolic method (PPM). The dark matter (DM) evolution is handled by the  particle-mesh technique commonly used in the Eulerian codes.

Our simulations start with cosmological initial conditions at z=100 and are generated from the "inits" package available with enzo. The parameters from the WMAP seven years data release are used to generate the initial conditions \citep{Jarosik2011}. The computational volume has a comoving size of 1 Mpc/h and periodic boundaries both for hydrodynamics and gravity. We  first run simulations with a uniform grid resolution of $\rm128^3$ cells (also $\rm 128^3$ DM particles) and select the most massive halos  forming in the  box at z=15.  The simulations are restarted with nested grid initial conditions,  one top and two nested refinement levels each with a grid resolution of $\rm 128^3$ cells. We employ  in total 5767168 DM particles to solve the dark matter dynamics which results in DM resolution of about $\rm 600~M_{\odot}$. In addition to this, we add 18 dynamical refinement levels during the course of simulations which provide an effective resolution of  about 1000 AU in the central 62 kpc comoving region. Our refinement criteria is based on  the gas density, the particle mass resolution and a  fixed resolution of 32 cells per Jeans length.  The cells are flagged for refinement if  their density is greater than four times the cosmic mean or the particle density is 0.0625 times  $ \rho_{DM}r^{\ell \alpha}$, here $r$ is the refinement factor (i.e. 2), $\ell$ is the refinement level, and $\alpha =-0.3$. To avoid spurious numerical affects, DM particles are smoothed at 12th refinement level, which corresponds to a resolution of about 2.7 pc in comoving units. 

\begin{figure*}
\includegraphics[scale=0.7]{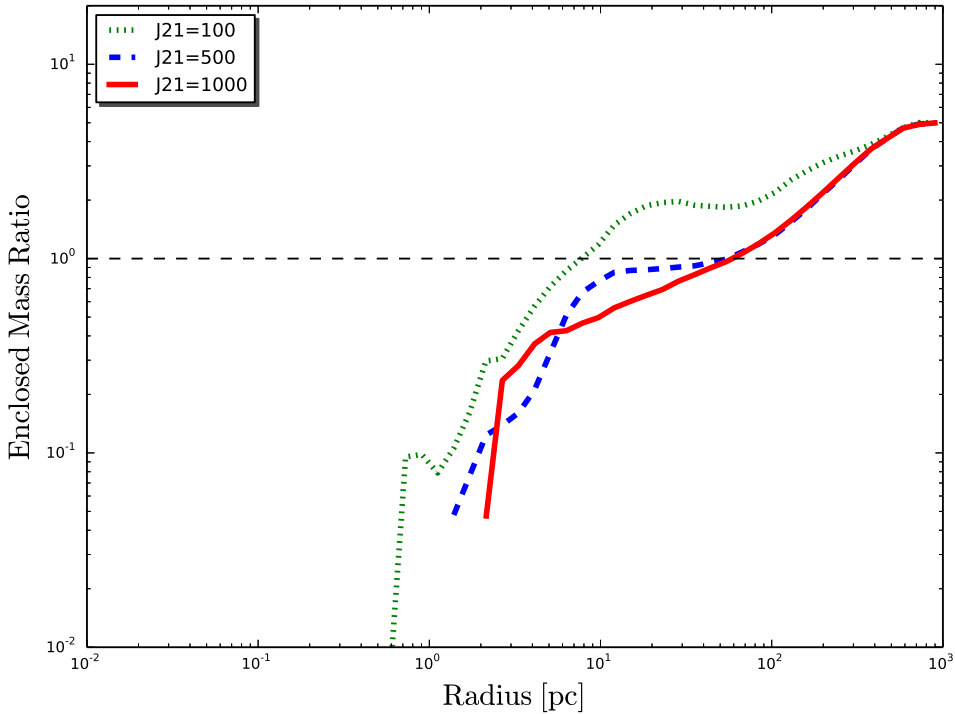} 
\caption{The ratio of enclosed DM mass to the enclosed baryonic mass is shown here for halo A, as a representative case.}
\label{fig00}
\end{figure*}

This technique of smoothing DM particles has been employed in numerous high resolution studies \citep{Wise2008, Turk2012,Latif2013c}. However, we note that \cite{Regan2015} have recently found that coarse DM resolution leads to spurious numerical effects such as artificial heating of the gas. This effect is stronger when no smoothing is applied to DM particles, and even smoothing is not sufficient in some cases, e.g., when the DM mass resolution is above 800 $\rm M_{\odot}$ for their cases with lower background UV flux. They suggest that to obtain numerically converged results the baryonic mass within the  core should be approximately 100 times the DM mass. To asses whether the baryonic core is properly resolved in our simulations, we show the ratio of enclosed DM to baryonic mass in figure \ref{fig00} for halo A, as a representative case.  Our results show that the DM mass dominates above 100 pc for $\rm J_{21}$= 500 and 1000 cases while for $\rm J_{21}$=100 it extends down to 10 pc. Below these scales, the ratio of enclosed masses ($\rm M_{DM}/M_{baryon}$) declines below unity for all BUV fluxes, with the same behaviour as in the  well resolved cases in \cite{Regan2015}. This shows that  at small radii the  baryonic core mass is about 50-100 times the  DM mass in our simulations, it is sufficiently resolved and not affected by  spurious DM resolution. Similar trend is observed for other two halos.

To solve the thermal evolution of the gas along with cosmological simulations, we employ the KROME package  \citep{Grassi2014}. Our chemical model is described in detail in \cite{Latif2015a} and  here we briefly summarise its main features. The rate equations of $\rm H$, $\rm H^{+}$, $\rm He$, $\rm He^{+}$,~$\rm He^{++}$, $\rm e^{-}$,~$\rm H^{-}$,~$\rm H_{2}$,~$\rm H_{2}^{+}$ are self-consistently solved with the hydrodynamics.  We employ a uniform isotropic background ultraviolet (BUV) flux with $\rm T_{rad}=2 \times 10^4$ K emitted by Pop II stars in units of $\rm J_{21}=10^{-21} ~erg/cm^2/s/Hz/sr$ and turn it on at z=30. Our model includes $\rm H_2$ formation, $\rm H_2$ photo-dissociation, $\rm H^-$ photo-detachment,  $\rm H_2$ collisional dissociation and the $\rm H_2$ self-shielding fitting formula of \cite{WolocottGreen2011}. We  include various cooling/heating mechanisms such as molecular hydrogen line cooling, cooling due to the collisional-induced emission, cooling/heating due to the three-body reactions, chemical cooling/heating and atomic line cooling (i.e.  cooling due to collisional excitation, collisional ionisation,  radiative recombination). Recently, \cite{Glover2015a} identified the key chemical reactions for computing $J_{crit}$, and provided a reduced chemical network for this purpose. Subsequently, \cite{Glover2015b} investigated the  impact of  uncertainties in the chemical reaction rates, comparing their results with several chemical networks employed in literature, including the KROME package adopted here.  In one-zone calculations, \cite{Glover2015b}  find that $J_{crit}$ is at most a factor of a few higher than when using the KROME. Therefore, our results provide a lower limit on the estimates of $J_{crit}$. However, we expect this difference to be further reduced in 3D simulations, as one-zone calculations often underestimate the value of $J_{crit}$ \citep{Latif2014UV,Latif2015a}.

\subsection{Sink Particles}
The simulations are evolved beyond the initial collapse by employing sink particles which represent the gravitationally bound objects. Such approach has been  successfully used both in smoothed particle hydrodynamics (SPH) and AMR codes \citep{Bate1995,Krumholz2004,Federrath2010,Latif2013d}.  The general criteria used to create sink particles include assuming a density threshold, requiring the highest level of refinement, having converging flow, or considering the local minimum of gravitational potential and/or a gravitationally bound region.  In the literature, various combinations of these approaches are used and a detailed comparison of these methods was made by \cite{Federrath2010}. They found that a simple density threshold criterion to create sink particles  is insufficient in the regime of supersonic turbulence because of local density enhancements. Recently, \cite{Bleuler2014} have presented a new method for sink particle creation in the Ramses code by employing a clump finding algorithm to more accurately locate the sites of  sink formation by taking into account the tidal effects of surrounding gas distribution.



We here employ the algorithm of \cite{Wang2010}  both for the creation of sinks, and their subsequent growth by accretion. Sinks are created only in the grid cells at the maximum refinement level, where, additionally, the Truelove criterion is violated, i.e., where the density exceeds the Jeans density, and in a region where the flow is convergent. Sinks are allowed to merge if they are formed within the accretion radius, i.e., the Jeans length in our case. Sinks are assigned masses so that a  cell becomes Jeans stable after the subtraction of the sink mass. Sink velocities are computed  based on momentum conservation. The works of \cite{Federrath2010} and \cite{Bleuler2014} suggest that this method of sink creation may create more sink particles. However, this may happen only in the regime where efficient fragmentation is expected, in the presence of supersonic turbulence or efficient cooling.  We do not impose any accretion criterion such as the Bondi-Hoyle  but  allow the creation of additional sinks  and let them merge within the accretion radius. We do not include any radiative feedback. If one assumes that sinks may represent  protostars, the stellar evolution calculations of \cite{Schleicher13} and \cite{Hosokawa2013} show that  the ionising  UV stellar feedback from the stars forming via rapid accretion, i.e., $\rm \geq 0.1 ~M_{\odot}/yr$ remains very low even after they reach $\rm 10^{4}~M_{\odot}$. Therefore, we expect that feedback  is negligible as long as accretion rates are $\rm \geq 0.1 ~M_{\odot}/yr$.

\begin{figure*}
\hspace{-6.0cm}
\centering
\begin{tabular}{c}
\begin{minipage}{6cm}
\hspace{-2cm}
\includegraphics[scale=0.2]{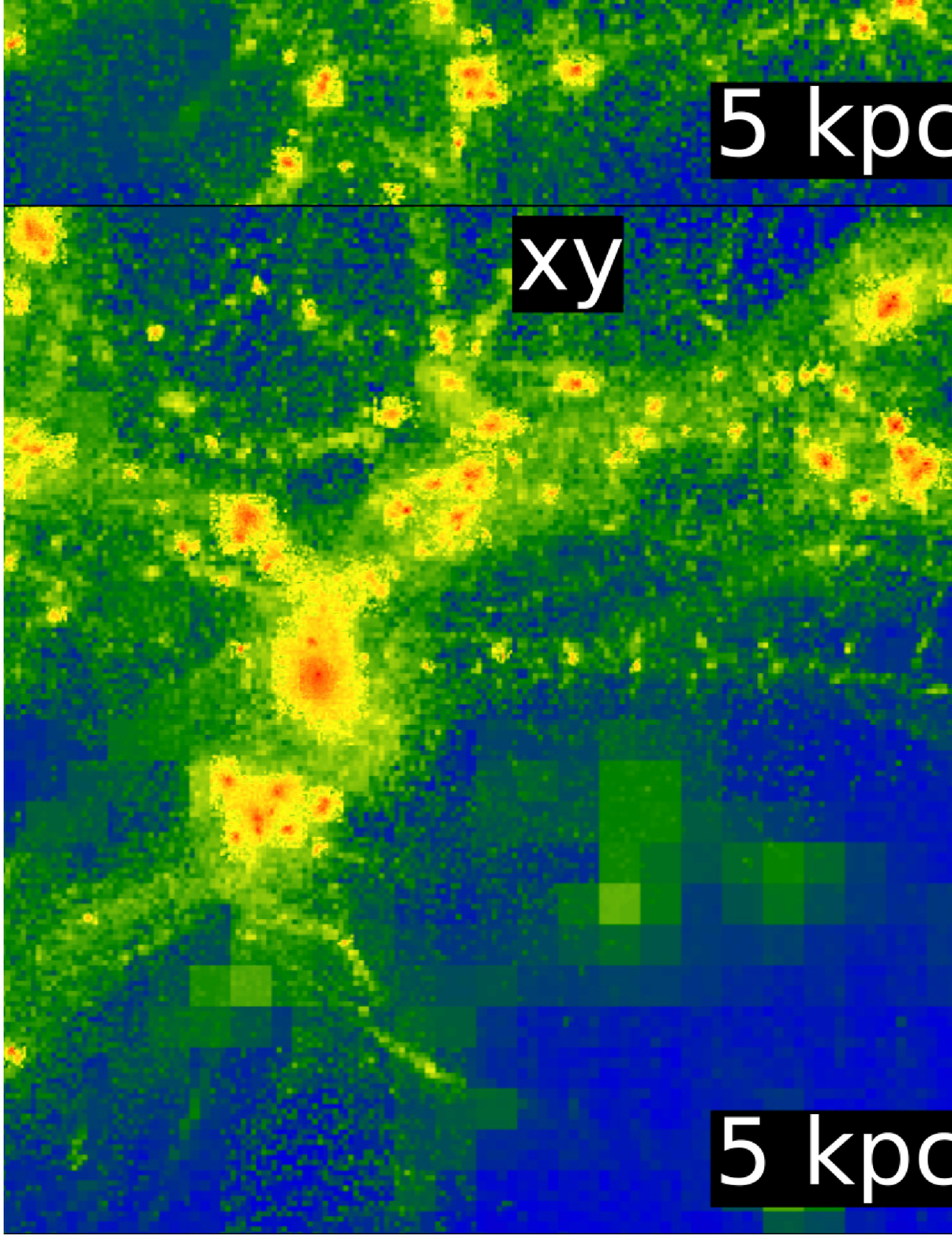}
\end{minipage} 
\end{tabular}
\caption{Average total (gas + DM) density  for  halo A illuminated by a BUV flux $\rm J_{21}=100$. The  columns show the  structure of the halo at various cosmic times and the rows  the  total density  averaged along the x, y \& z axis. The width of the density projections is given in  physical units and the time in Myr after the Big Bang. In the middle column, the white circles highlight the main halo and the halo merging with it.}
\label{fig0}
\end{figure*}

\section{Results}
 We perform three dimensional cosmological simulations for  three distinct massive primordial halos of a few $\rm 10^7~M_{\odot}$  for various strengths of the BUV flux. The properties of the halos such as their masses, collapse redshifts, spins and the critical strength of UV flux are listed in table \ref{table1}.  In the following, we  discuss the structural properties of  the halos and their temporal evolution.  We also study the thermodynamical and physical properties of the halos. We distinguish here between the mass inflow rate, provided by the halo, and the accretion rate that occurs on a putative protostar.  The mass inflow rates  are computed for the each intensity of the BUV flux and the mass accretion is followed by employing sink particles for a subset of halos.

  \begin{table}.
\begin{center}
\caption{Eigenvalues of the inertia tensor for halo A.}
\begin{tabular}{ccccccc}
\hline
\hline

$\rm J_{21}$	&  $\rm a_{1}$  & $\rm a_{2}$ & $\rm a_3$ \\

\hline                                                          \\
$J_{21}=100$  &7.9  &7.36  &4.37 \\
$J_{21}=500$  &3.0  &2.17  &2.14 \\
$J_{21}=1000$ &3.5  &3.4  &2.9 \\
$J_{21}=40000$  &7.7  &7.5  &6.7 \\

\hline
\end{tabular}
\label{tab2}
\end{center}
\end{table}

\subsection{ Structural properties of the halos}
According to our understanding of structure formation, the quantum fluctuations in the gravitational potential induced density perturbations. These perturbations collapsed via gravitational instabilities and merged with each other to form halos. Baryons fall in the DM potentials, get shock heated and this process continues until the virialization of the halo. We see that minihalos of a few times times $\rm 10^5 ~M_{\odot}$ are formed at about z=20 in our simulations.  Due to the presence of a BUV flux, these minihalos are unable to collapse via molecular hydrogen cooling. They continue to grow via  mergers and accretion from the cosmic web until their virial temperature  exceeds $\rm 10^4$ K. In figure \ref{fig0}, we show the structure of halo A  at various cosmic times. This  halo is formed at the intersection of filaments and is mainly fed by three main streams. The mass of the halo $\rm \sim$ 300 Myrs ($z \sim 13$) after the Big Bang is $\rm 1.4 \times 10^7~M_{\odot}$ and it undergoes a major merger with a halo of $\rm 5 \times 10^6~M_{\odot}$ (i.e., mass ratio $\sim$1:2) about 100 Myrs later. The halo virializes at $ \rm \sim z = 11$, when its mass is well above the atomic cooling limit.  In the meantime,  it is continuously fed by  gas flowing  from the filaments and by minor mergers. 

Halo B goes through many minor mergers with mass ratios of about 1:10 and is also accreting  gas via cosmic streams. Eventually, about 300 Myrs after the Big Bang,  the halo becomes virialized, and has a mass of about $\rm 2.7 \times 10^7~M_{\odot}$.  This halo has a very different merger history from  halo A as it has not experienced any major merger recently and is assembled  $\sim$130 Myr earlier than halo A.  Similar behaviour is observed for all BUV flux strengths as the incident flux does not impact the large scale properties of the halos. Halo C experienced a major merger about 350 Myr after the Big Bang,  similar to the history of  halo A. The mass ratios of the merging halos is about 1:2 and the halo gets virialized about  50 million years later. 


Due to the presence of angular momentum in gas and halos, a central rotationally supported structure is expected to form as a consequence of  gravitational collapse \citep{Oh2002,Volonteri2005}. However, as discussed by \cite{Choi2014}, the triaxiality of dark matter halos \citep{Allgood2006}  exerts gravitational torques and helps in transferring angular momentum via bars in bars instabilities, thus reducing rotational support compared to an axisymmetric case. In order to study the properties of the central regions, specifically their level of rotational support, we have computed the ratio between the rotational velocity,$V_{rot}$,  and the velocity dispersion, $\sigma$. This ratio, $V_{rot}/ \sigma$,  provides us the estimate of the disk scale height as  $\rm H/R \sim$ $\sigma/{\rm V_{rot}}$ where $H$ is the disk height, and $R$ is the disk radius. Our estimates of $V_{rot}/ \sigma$ for  halo A are shown in the bottom-right panels of figures \ref{fig3}, \ref{fig4}, \ref{fig5} and \ref{fig6} for $\rm J_{21}$=100, 500, 1000 and $\rm 4 \times 10^4$.  $V_{rot}/ \sigma$ is $\sim$ 6 in the central 10 pc of the  halo for $\rm J_{21}$=100. Rotational support is higher at lower  $\rm J_{21}$ values, as cooling in the core of the halo is efficient, because the fraction of molecular hydrogen (shown in the top-left panels) is higher.  $V_{rot}/ \sigma$ in fact becomes smaller for higher fluxes and approaches unity for the isothermal case. We estimate that the aspect ratio (i.e. H/R) is $\rm \sim 0.25$ for $\rm J_{21}=100$ and  increases up to $0.5$ for $\rm J_{21}=1000$. The aspect ratio is almost unity for the isothermal case, i.e., for $\rm J_{21}=4 \times 10^4$. This suggests that a fat disk for all radiation field strengths is formed in the halo with aspect ratio higher than 0.1.  Similar behaviour is observed for halo C (not shown). The value of $V_{rot}/ \sigma$ is higher in the case of halo B (see figures, \ref{fig7}, \ref{fig8} and \ref{fig9}) in general, but it decreases as collapse approaches the isothermal state \citep[cf.][]{Oh2002}. It indicates that the scale height of the disk  varies from halo to halo.

We also calculated the components of the inertia tensor in the following way:
\begin{equation}
I_{xx} = \sum_{j=1, N} m_j (y_{j}^2 + z_{j}^2)
\label{eq1}
\end{equation}
\begin{equation}
I_{xy} = - \sum_{j=1, N} m_j x_{i}y_{j} .
\label{eq1}
\end{equation}
 The other components of the inertia tensor can be computed in a similar fashion. The diagonal elements of the inertia tensor are known as the moments of inertia. The eigenvalues a$_1$, a$_2$ and a$_3$ of the inertia tensor for  halo A are listed in table \ref{tab2}  and are computed at the same time at which $V_{rot}/ \sigma$  is measured (i.e., last time evolution in figures \ref{fig3}, \ref{fig4}, \ref{fig5} and \ref{fig6}). For $\rm J_{21}=100$,  the eigenvalues of two components are almost equal but the value for the perpendicular component is 1.8 times  higher than the other two. The expected value for the perpendicular component of a thin disk is  twice the values in the disk plane in agreement with our results. Similarly, the eigenvalues of two components are almost equal for $\rm J_{21}=500$ , 1000 and 40000 while the value of the perpendicular component is 1.4, 1.2 and 1.1 times the values in the disk plane respectively.  This suggests that the disk is rather thin for $\rm J_{21}=100$ and becomes thicker  at higher fluxes.  Similar trends are observed for the other two halos.

We compute the rotational velocity by taking the ratio of the specific angular momentum  $\rm \vec{\ell}$ and the position vector $\rm \vec{r}$ ($V_{rot} \sim \vert \vec{\ell} \vert /  \vert \vec{r} \vert $, the definition used in our  previous analysis) and compare it with the definition  $\rm \vert \vec{\ell} \vert /  \vert \vec{a_{1}} \vert$  given in the equation 13 of \cite{Regan09}. These two definitions differ only by a factor of 1.3. \cite{Regan09} found similar differences, as expected in turbulent collapsing gas where rotation is not always well defined.  We show the  ratio of rotational to circular velocity in figure \ref{figvratio}. For the isothermal collapse case the ratio of $\rm V_{rot}/V_{cir}$ remains below unity except in the very centre (within the Jeans length) where we approach the resolution limit.  The increase in $\rm V_{rot}$ above the $\rm V_{circ}$ by a factor of five for the weaker flux cases  could be a consequence of various factors such as ellipsoidal rather than spherical geometry and large infall rates that keep the structure unsettled. This analysis confirms that  rotational support is more significant in halos illuminated by low BUV fluxes  compared with halos collapsing isothermally. The higher rotational support in some halos can delay  collapse and may even halt it if  angular momentum is not transferred efficiently. In such a case, it may limit accretion onto the central object and  have important implications for the final masses of stars.

\subsection{Thermodynamical properties}

The thermal properties of  halo A for various strengths of BUV flux are shown in the figures \ref{fig3}, \ref{fig4}, \ref{fig5}  and  \ref{fig6}. The temperature of the halo is  a few thousand K when the halo mass is lower than the atomic cooling threshold. It increases through virialization and merger shocks until it reaches the atomic cooling limit.  We note that in the presence of  a BUV flux, irrespective of its strength, the temperature of the gas in the halo is $\sim$ 8000 K after its virialization at scales larger than 10 pc.  This is due to the lack of molecular hydrogen cooling as its abundance remains low, i.e. $\rm < 10^{-6}$. At smaller scales, $\rm H_2$ molecules self-shield  from the radiation depending on the strength of the ambient UV flux.  For radiation field strengths $\rm < J_{21}=10^4$, $\rm H_2$ self-shielding becomes effective due to its ($\rm H_2$) enhanced abundance and consequently the gas temperature decreases  to a few hundred K within the central 10 parsec. To further illustrate this effect, we show the $\rm H_2$ column density in figure \ref{figH2col} for various BUV fluxes for halo A.  As expected, the $\rm H_2$ column density becomes larger than $\rm 10^{16}~cm^{-2}$ for weaker fluxes. Consequently, $\rm H_2$ self-shielding becomes effective and $\rm H_2$ cooling occurs in the centre. On the other hand, for the strongest flux (i.e., $\rm J_{21}=4 \times 10^4$) the $\rm H_2$ column density remains lower than $\rm 10^{15}~cm^{-2}$, self-shielding  remains ineffective and as a result the collapse is isothermal. We define the transition from atomic to molecular hydrogen cooling at $\rm H_2$ abundance above $\rm 10^{-4}$, which occurs between 1-10 pc and  is delayed by a few Myrs  for the $ \rm J_{21}=1000$ case in comparison with  $ \rm J_{21}=100$. The thermal evolution of  halo B for $\rm J_{21} =$ 100, 500 and 1000 is also very similar as shown in figures \ref{fig7}, \ref{fig8}  and \ref{fig9}. Halo C (not shown) has also a very consistent behaviour. Their cores are cooled by molecular hydrogen with central temperatures of  a few hundred K and  have higher temperatures at larger radii.  For the weakest flux cases, $\rm H_2$ cooling becomes important earlier than for the stronger flux cases. Variations are observed from  halo to halo, e.g., in halo C the transition from atomic to molecular state occurs at 0.4 pc\footnote{See  \cite{Latif2014UV} and \cite{Latif2015a} where abundances of various chemical species are compared for the halos studied here.}.  Similarly, variations from halo to halo are also observed \citep [e.g.,] []{Shang2010}.

In the case of the strongest flux, $\rm J_{21} = 4 \times 10^4$, the formation of molecular hydrogen remains suppressed throughout and the halo collapses isothermally with a central temperature  of  8000 K. The fraction of $\rm H_2$ in this case remains below $\rm10^{-6}$. Similar behaviour is observed for the other two halos for such high flux. Isothermal collapse occurs therefore only for significantly high values of $\rm J_{21}$, $\rm J_{21} > 10^4$.  The density inside the halo increases during the collapse and the peak density reached in the simulations is a  few times  $\rm 10^{-17}~ g/cm^{-3}$.  The density profile in halo A  follows approximately an $\rm  R^{-2}$ behaviour and becomes constant within the Jeans length. The small bumps in the profile indicate the presence of additional substructure in the halo. The amount of substructure in the halos reduces with increasing flux strength.  Similar features are observed for the other two halos. 

%
%
%

\subsection {Mass inflow rates and masses at initial stages}

One of the prime objectives of the present work is to compute the mass inflow rates in the simulated halos for different thermodynamical conditions and assess whether they are sufficient to build a supermassive star.  The latter requires mass accretion rates of about  $\rm \geq 0.01-0.1~ M_{\odot}/yr$ to rapidly form a massive object \citep{Begelman2010,Hosokawa2013,Schleicher13}. 

\begin{figure*}
\includegraphics[scale=0.7]{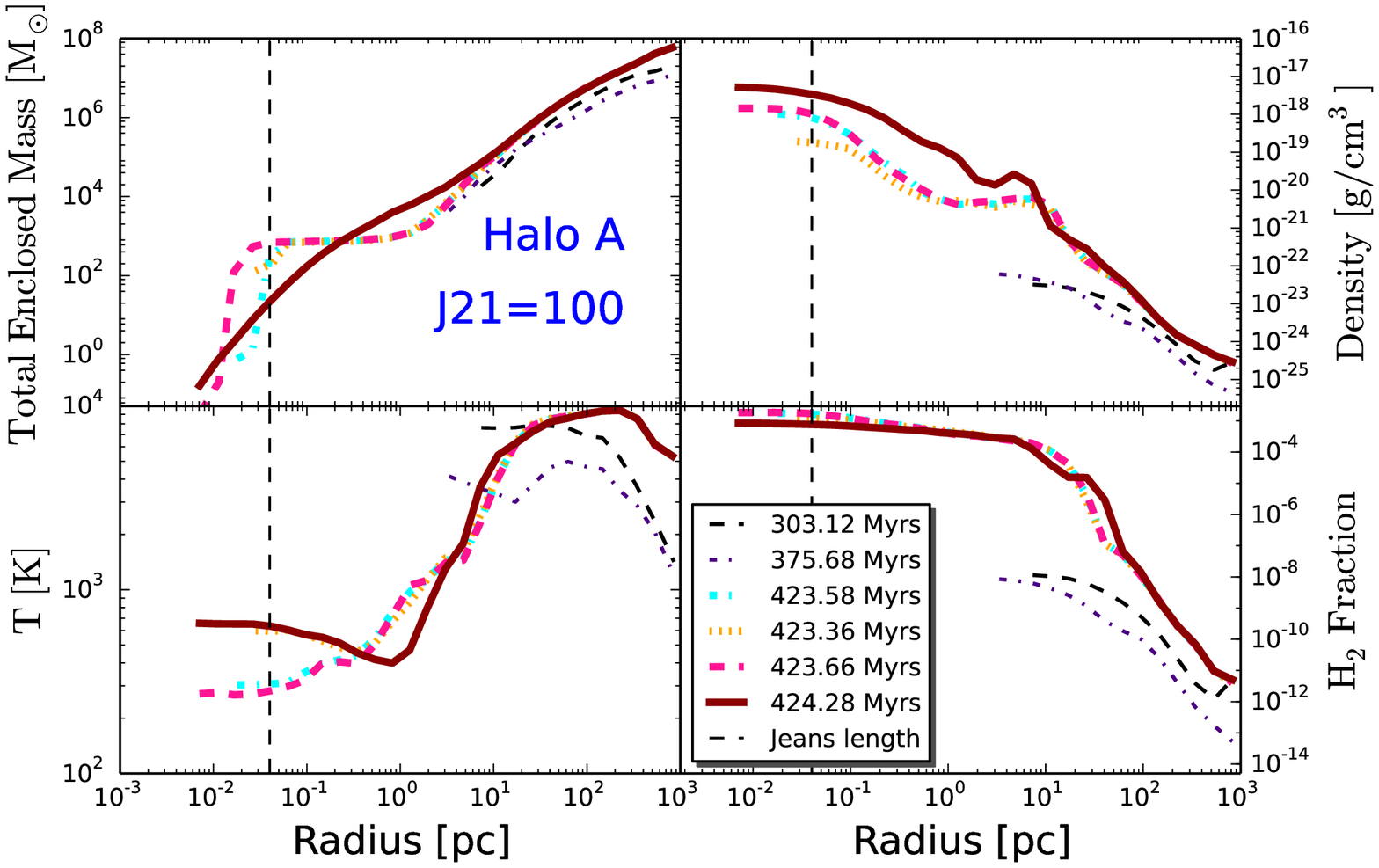}  \\
\includegraphics[scale=0.7]{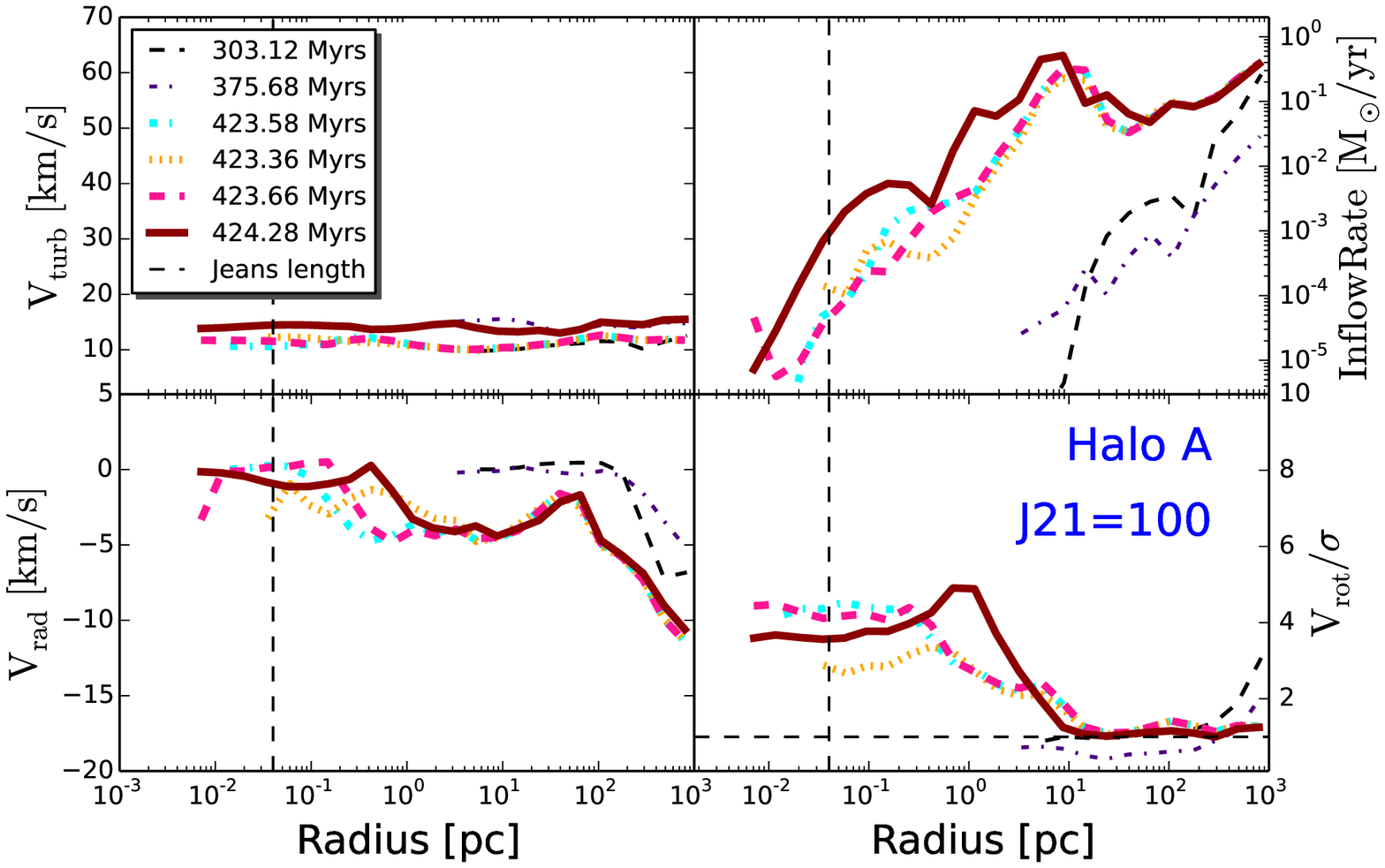} 
\caption{The thermodynamical and physical properties of halo A are shown for  $\rm J_{21}= 100$. The top panels show the time evolution of gas density, temperature, $\rm H_2$ fraction and of the total enclosed mass. The  time evolution of the inflow rates, V$_{rot}/ \sigma$, radial infall and turbulent velocities is shown in the bottom panels. The different line styles represent various cosmic times. The profiles are spherically averaged and radially binned. The vertical dashed black line shows the Jeans length and the horizontal dashed line represents V$_{rot}/ \sigma$ =1.}
\label{fig3}
\end{figure*}

\begin{figure*}
\includegraphics[scale=0.7]{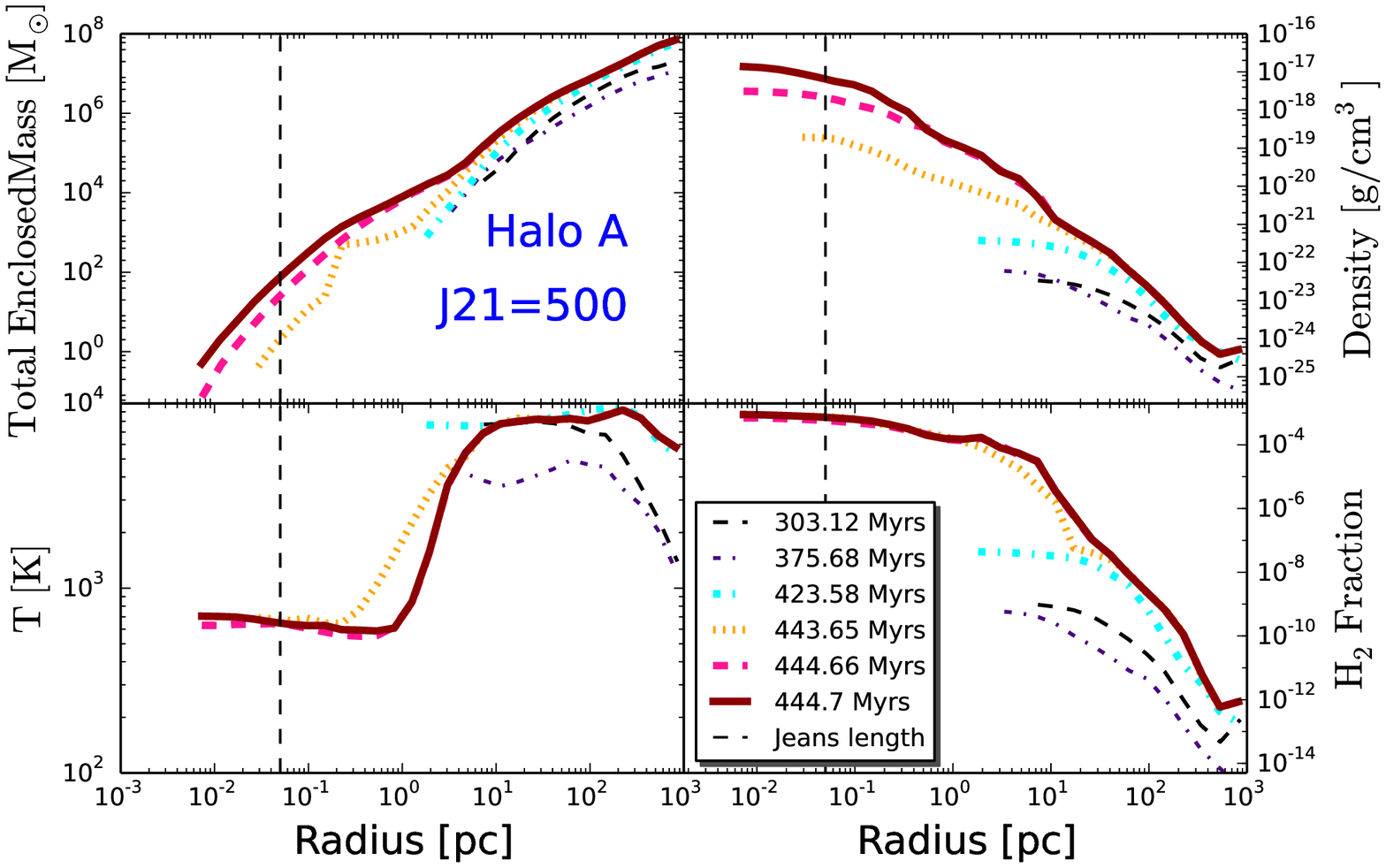} \\
\includegraphics[scale=0.7]{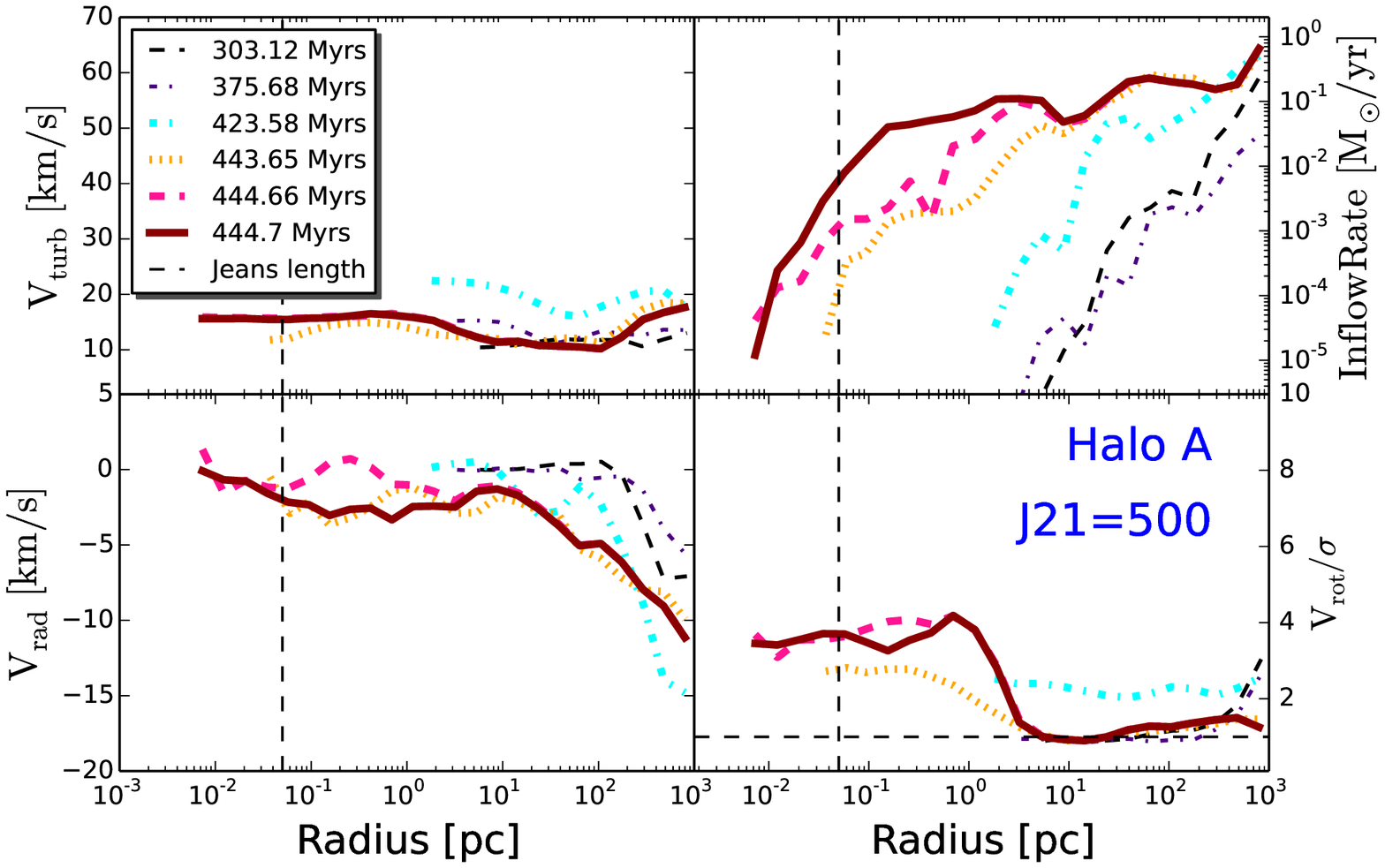} 
\caption{The thermodynamical and physical properties of halo A are shown for  $\rm J_{21}= 500$. The top panels show the time evolution of gas density, temperature, $\rm H_2$ fraction and of the total enclosed mass. The  time evolution of the inflow rates, V$_{rot}/ \sigma$, radial infall and turbulent velocities is shown in the bottom panels. The different line styles represent various cosmic times. The profiles are spherically averaged and radially binned. The vertical dashed black line shows the Jeans length and the horizontal dashed line represents V$_{rot}/ \sigma$ =1.}
\label{fig4}
\end{figure*}

\begin{figure*}
\includegraphics[scale=0.7]{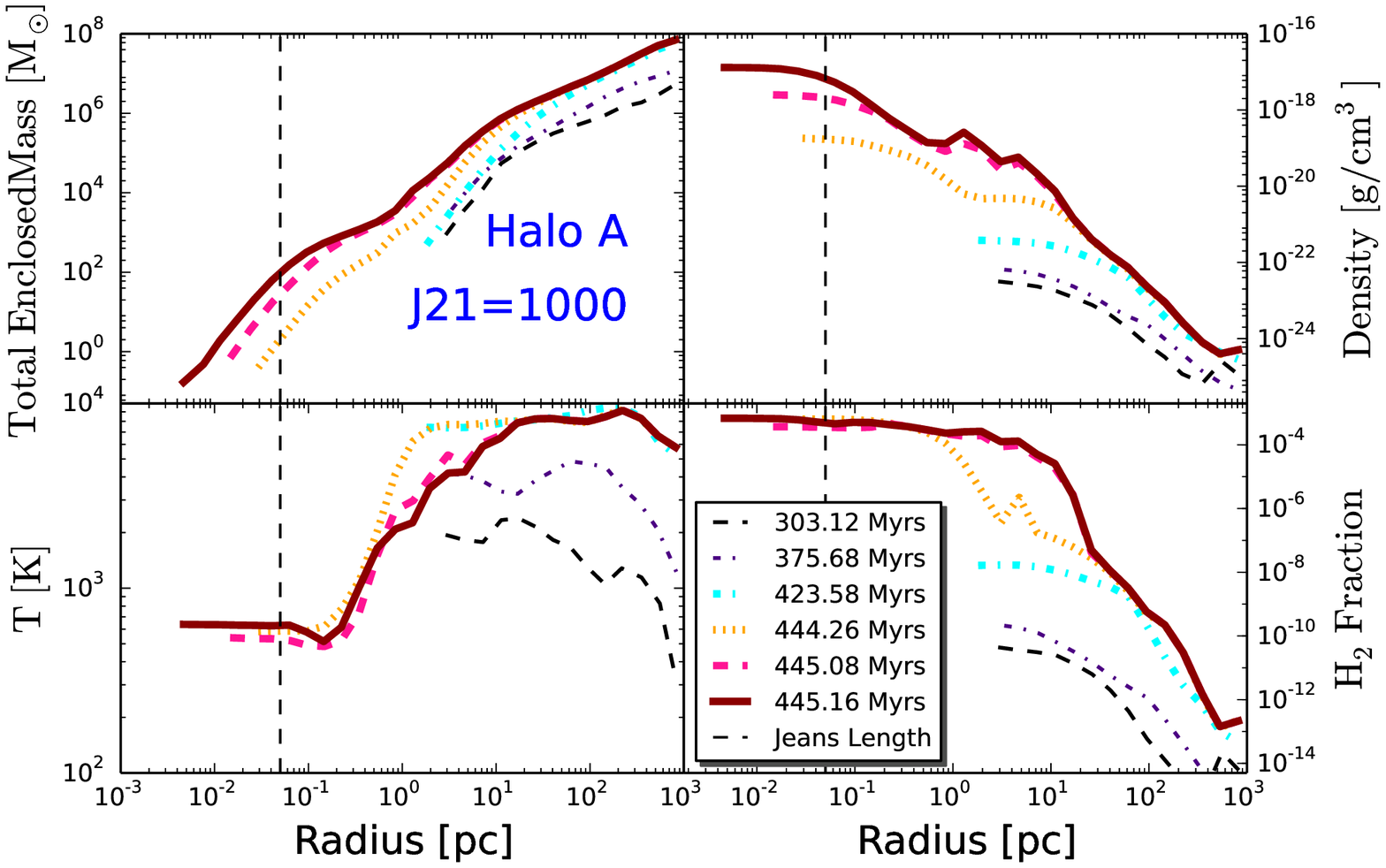} \\
\includegraphics[scale=0.7]{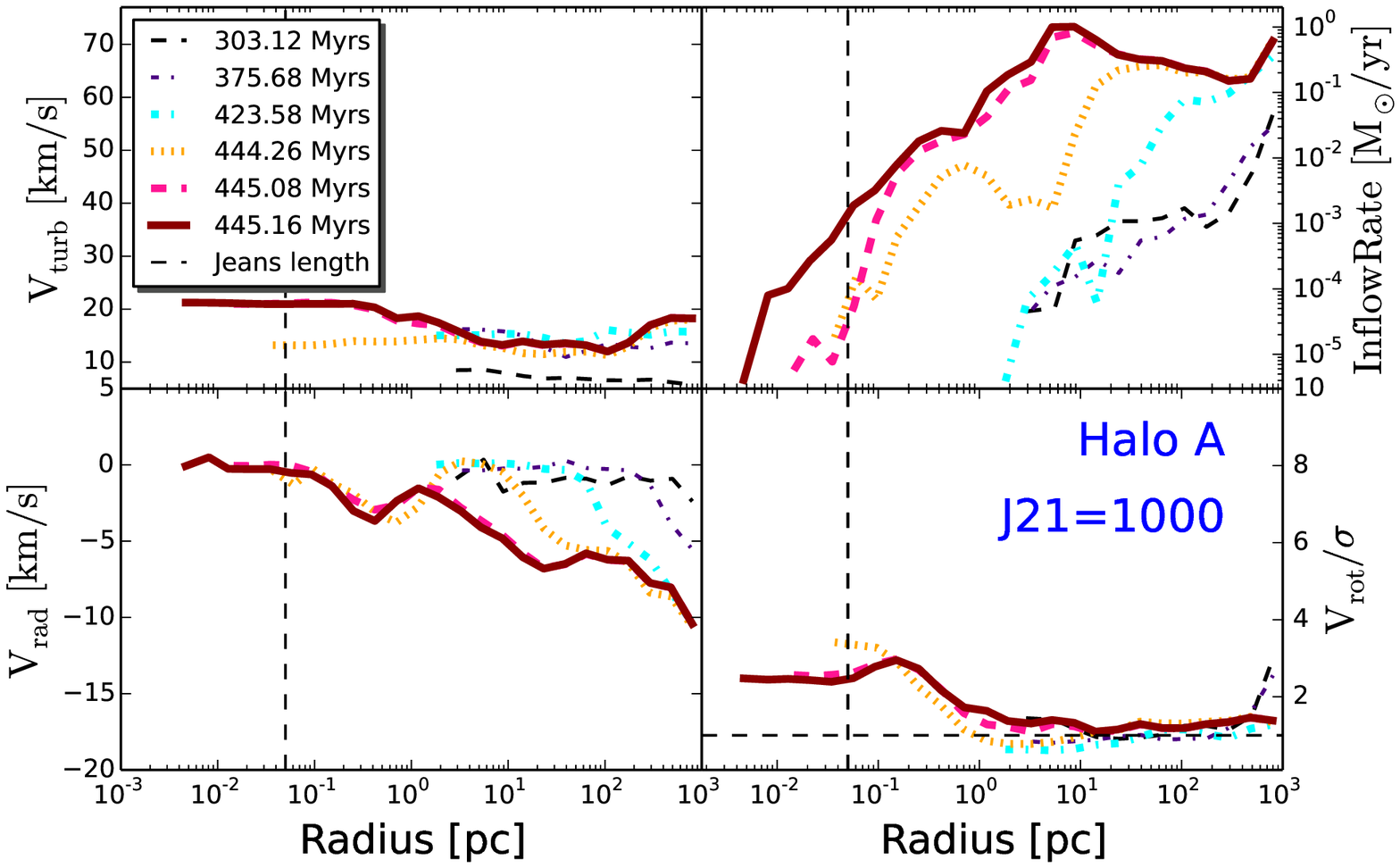} 
\caption{The thermodynamical and physical properties of the halo A are shown for  $\rm J_{21}= 1000$. The top panels show the time evolution of gas density, temperature, $\rm H_2$ fraction and of the total enclosed mass. The  time evolution of the inflow rates, V$_{rot}/ \sigma$, radial infall and turbulent velocities is shown in the bottom panels. The different line styles represent various cosmic times. The profiles are spherically averaged and radially binned. The vertical dashed black line shows the Jeans length and the horizontal dashed line represents V$_{rot}/ \sigma$ =1.}
\label{fig5}
\end{figure*}

\begin{figure*}
\includegraphics[scale=0.7]{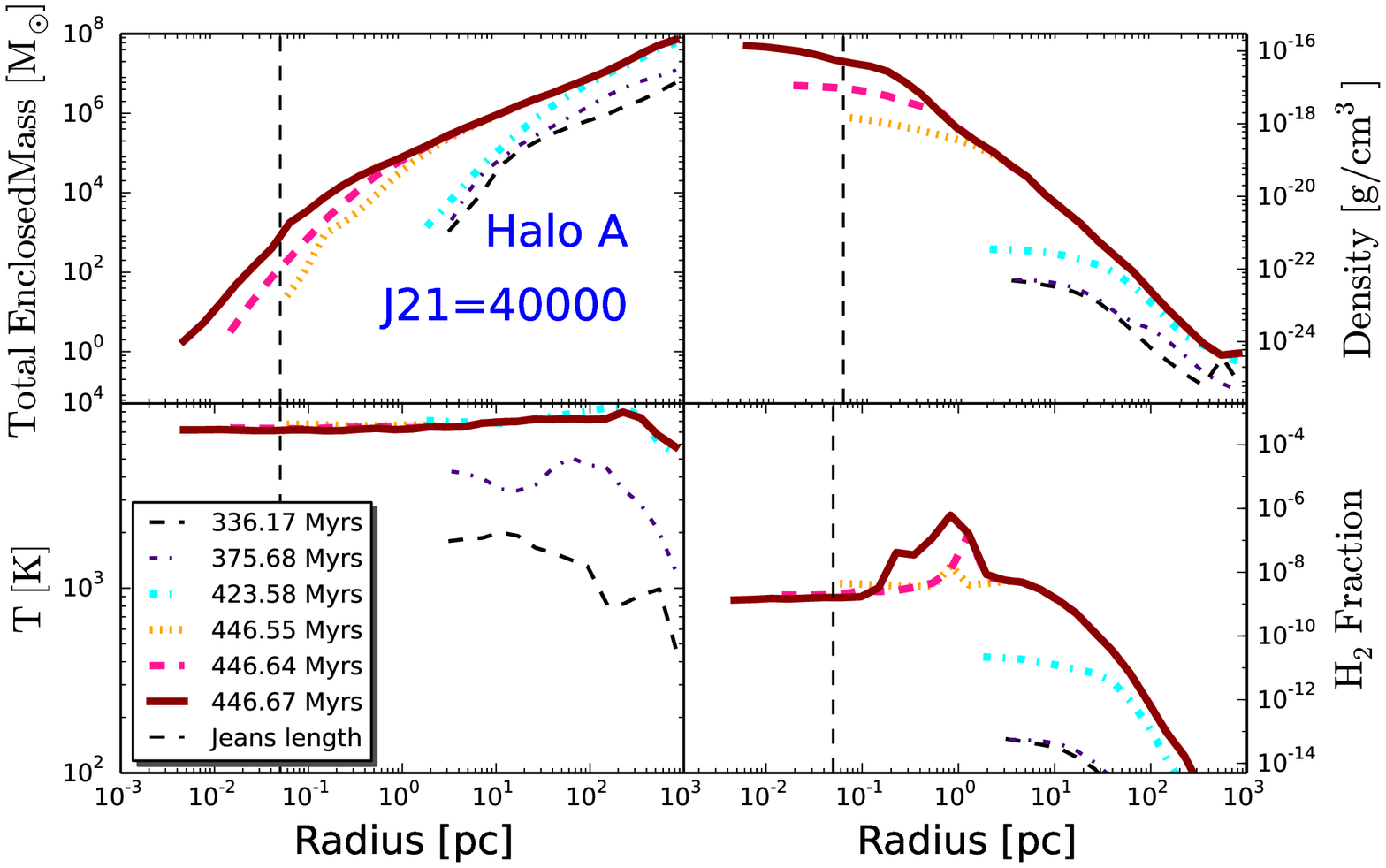} \\
\includegraphics[scale=0.7]{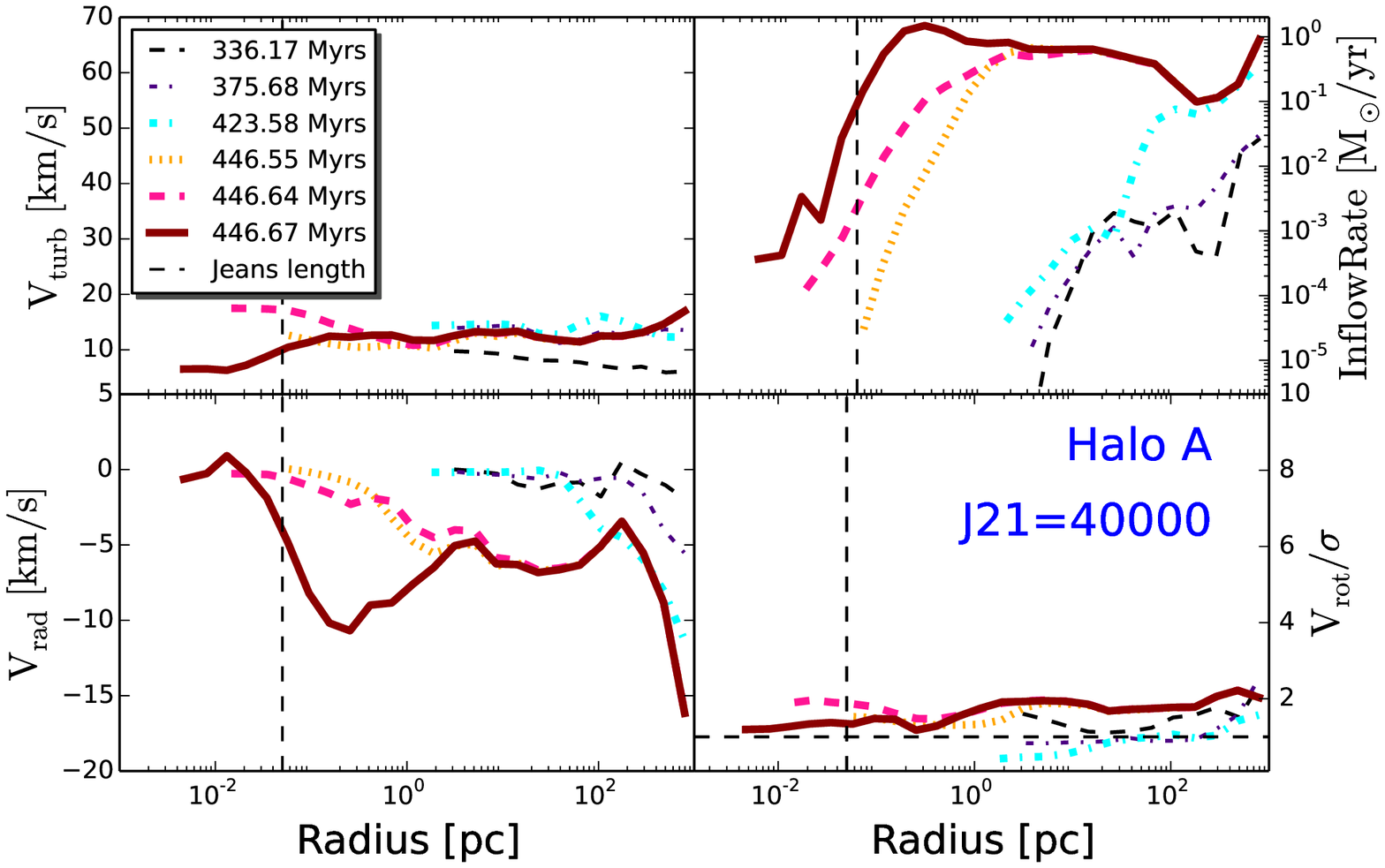} 
\caption{The thermodynamical and physical properties of the halo A are shown for  $\rm J_{21}= 4 \times 10^4$. The top panels show the time evolution of gas density, temperature, $\rm H_2$ fraction and of the total enclosed mass. The  time evolution of the inflow rates, V$_{rot}/ \sigma$, radial infall and turbulent velocities is shown in the bottom panels. The different line styles represent various cosmic times. The profiles are spherically averaged and radially binned. The vertical dashed black line shows the Jeans length and the horizontal dashed line represents V$_{rot}/ \sigma$ =1.}
\label{fig6}
\end{figure*}

The temporal evolution of the mass inflow rates  for  halo A in the presence of various BUV flux strengths is shown in figures \ref{fig3}, \ref{fig4}, \ref{fig5} and  \ref{fig6}. The mass inflow rate is about $\rm 0.001- 0.01~M_{\odot}/yr$  when the halo mass is $\rm \sim 10^6~ M_{\odot}$.  It increases up to  $\rm  \sim 0.1 ~M_{\odot}/yr$ when the halo mass is a few times $\rm 10^7~M_{\odot}$ for all the strengths of BUV flux  ($\rm J_{21}$= 100, 500 and 1000).  This is comparable to the theoretical  free-fall rate, i.e. $\rm \dot{M} \sim {c_s^3}/{G} \sim 0.1~M_{\odot}/yr \left( {T}/{8000~K}\right)^{3/2}$,  where $c_s$ is the thermal sound speed. The inflow rate tends to be  higher for $\rm J_{21}=1000$  at  all radii  compared to the $\rm J_{21}=100$ but differences are not significant. Similar values are found for $\rm J_{21}=500$. In general, the small fluctuations in the inflow rates at various radii are due to the different density structure and the intermittent flows present in the halo (note that these are spherical averages).

\begin{figure*}
\includegraphics[scale=0.7]{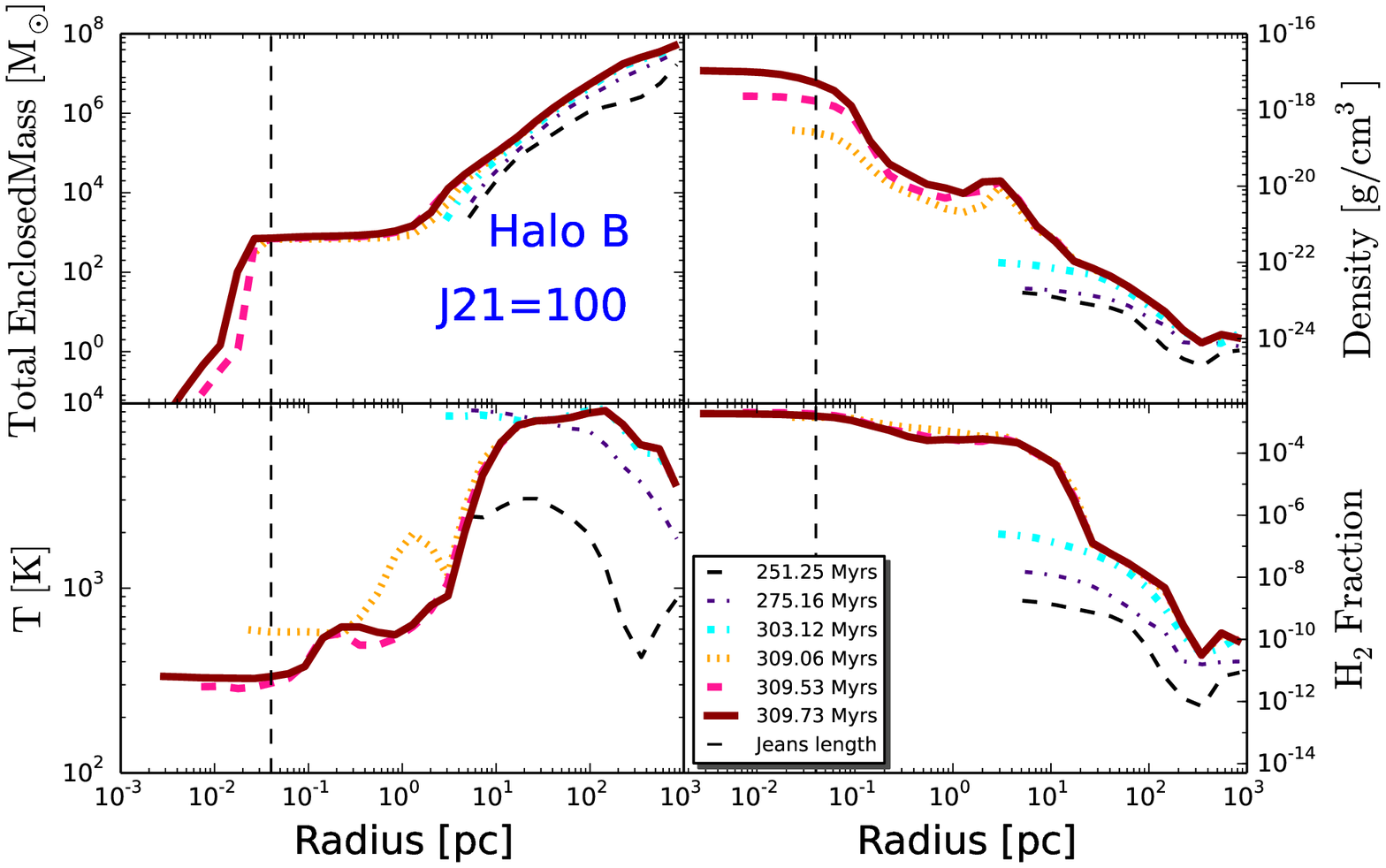} \\
\includegraphics[scale=0.7]{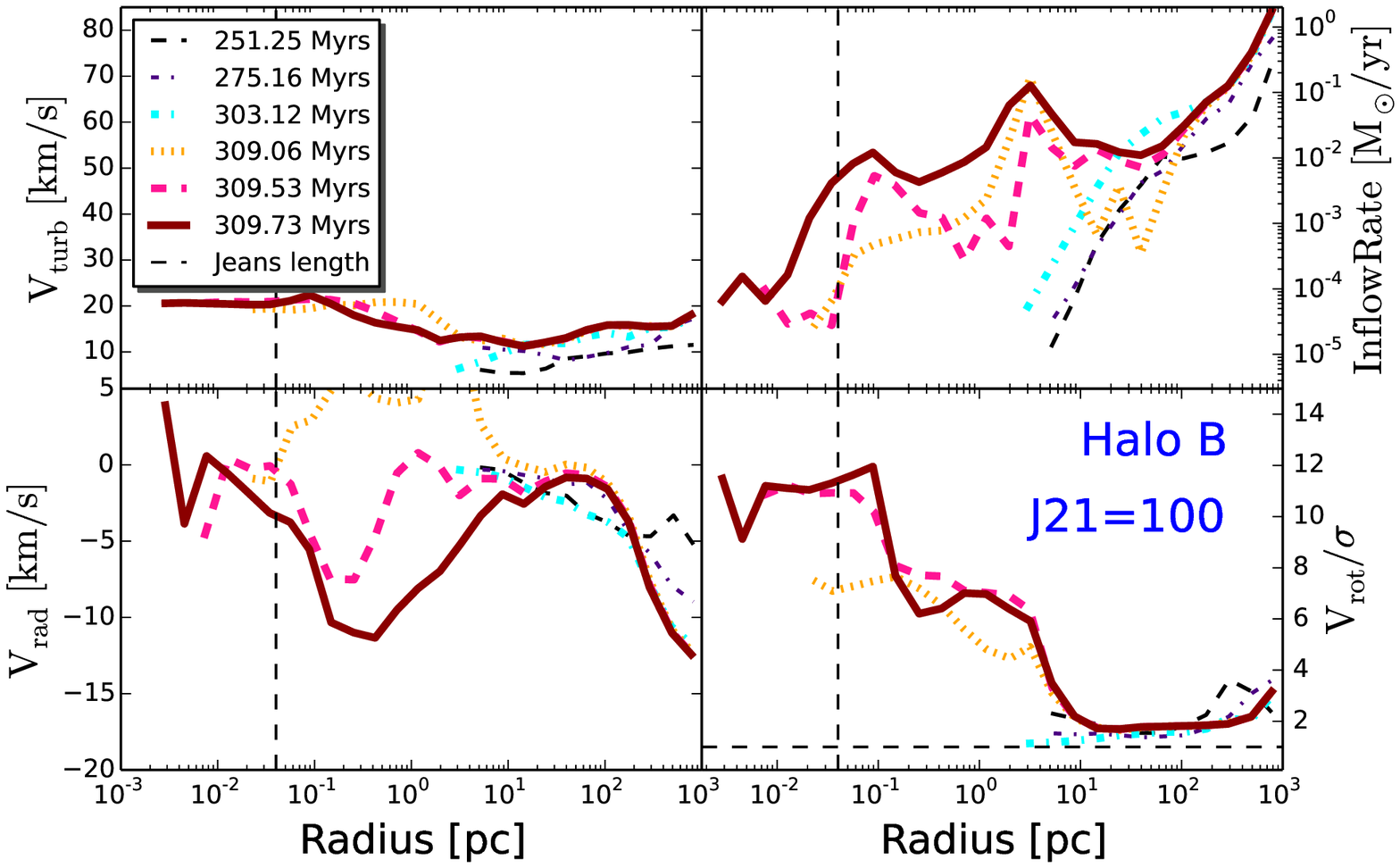} 
\caption{The thermodynamical and physical properties of the halo B are shown for  $\rm J_{21}= 100$. The top panels show the time evolution of gas density, temperature, $\rm H_2$ fraction and of the total enclosed mass. The  time evolution of the inflow rates, V$_{rot}/ \sigma$, radial infall and turbulent velocities is shown in the bottom panels. The different line styles represent various cosmic times. The profiles are spherically averaged and radially binned. The vertical dashed black line shows the Jeans length and the horizontal dashed line represents V$_{rot}/ \sigma$ =1.}
\label{fig7}
\end{figure*}

\begin{figure*}
\includegraphics[scale=0.7]{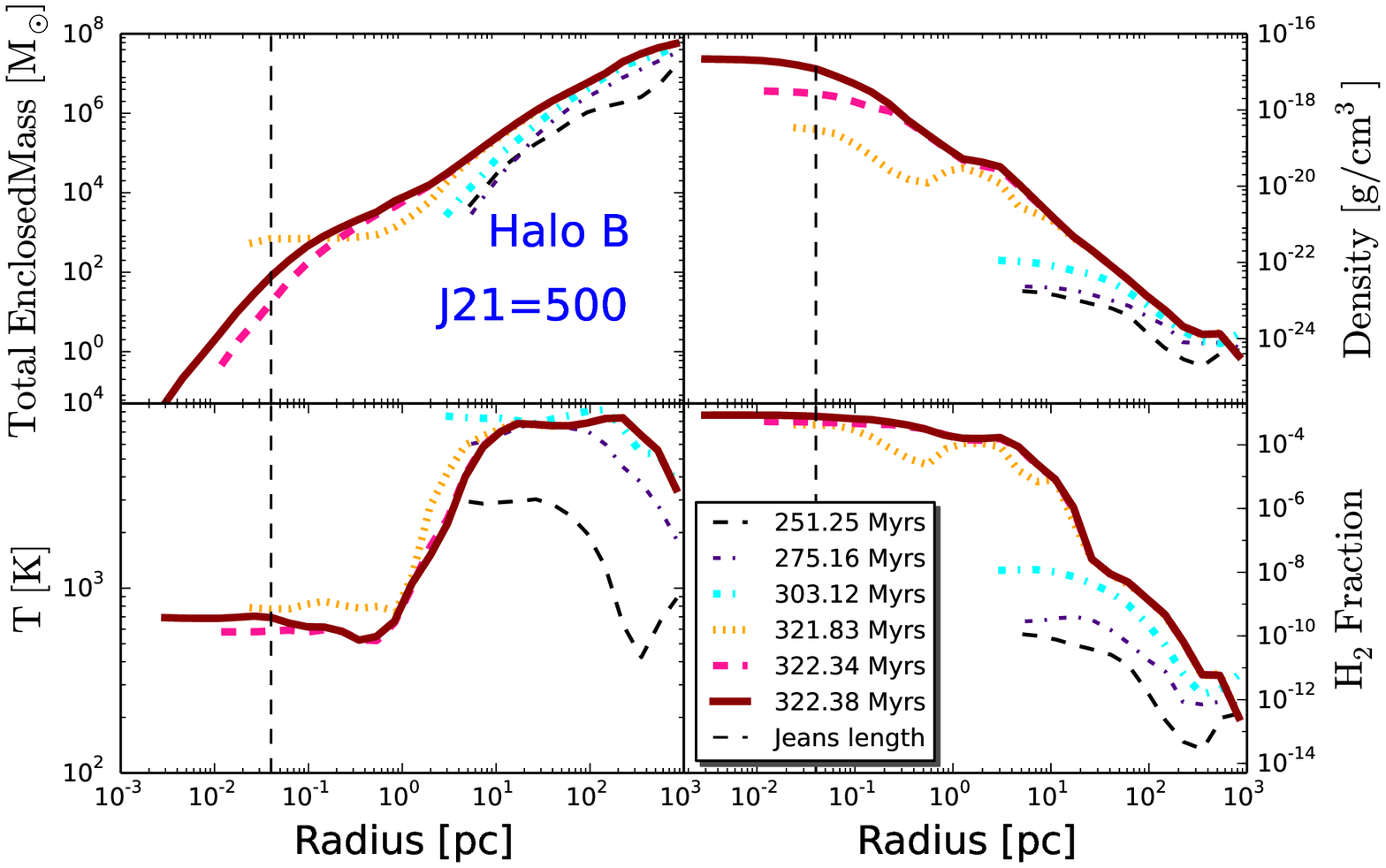} \\
\includegraphics[scale=0.7]{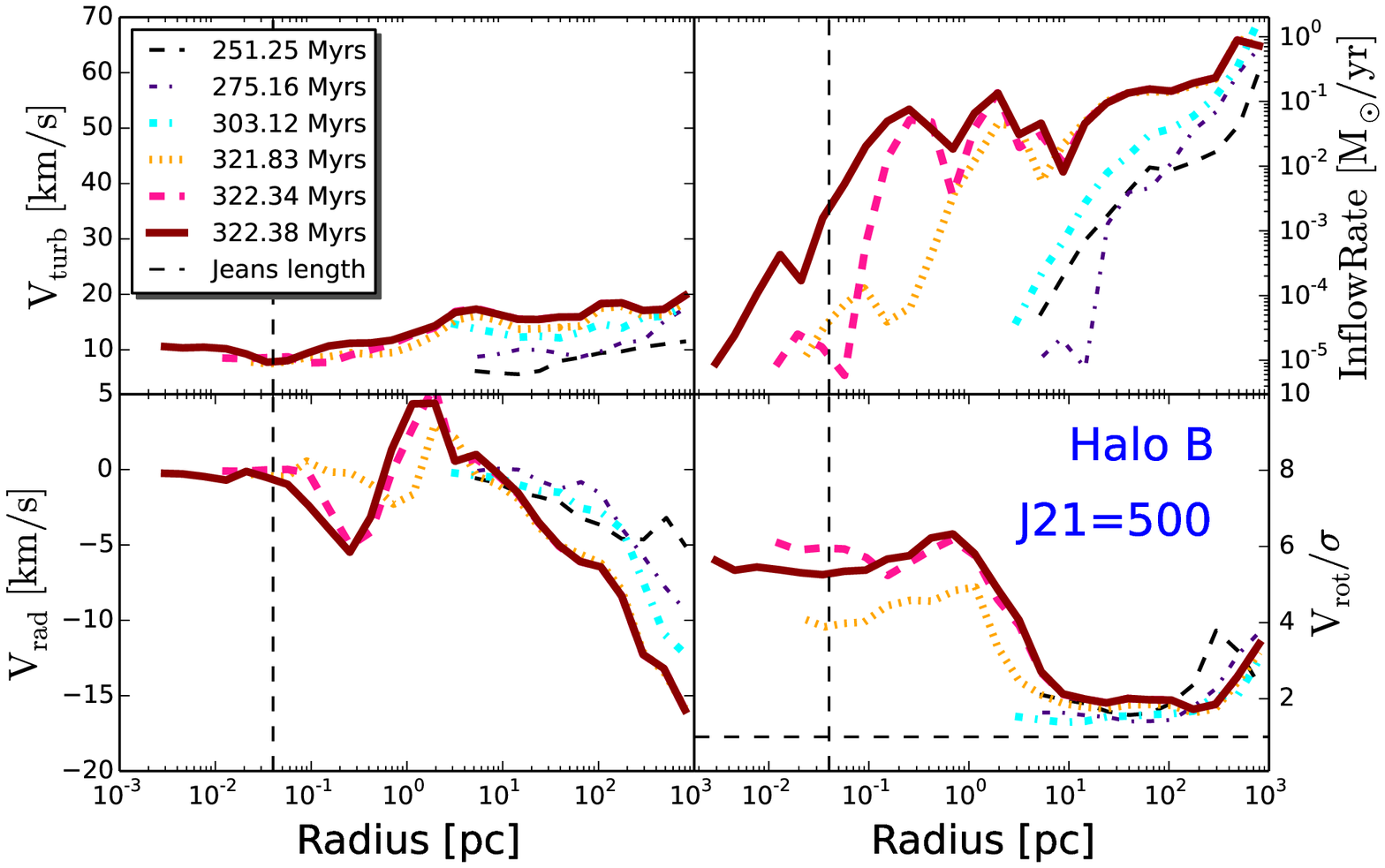} 
\caption{The thermodynamical and physical properties of the halo B are shown for  $\rm J_{21}= 500$. The top panels show the time evolution of gas density, temperature, $\rm H_2$ fraction and of the total enclosed mass. The  time evolution of the inflow rates, V$_{rot}/ \sigma$, radial infall and turbulent velocities is shown in the bottom panels. The different line styles represent various cosmic times. The profiles are spherically averaged and radially binned. The vertical dashed black line shows the Jeans length and the horizontal dashed line represents V$_{rot}/ \sigma$ =1.}
\label{fig8}
\end{figure*}

\begin{figure*}
\includegraphics[scale=0.7]{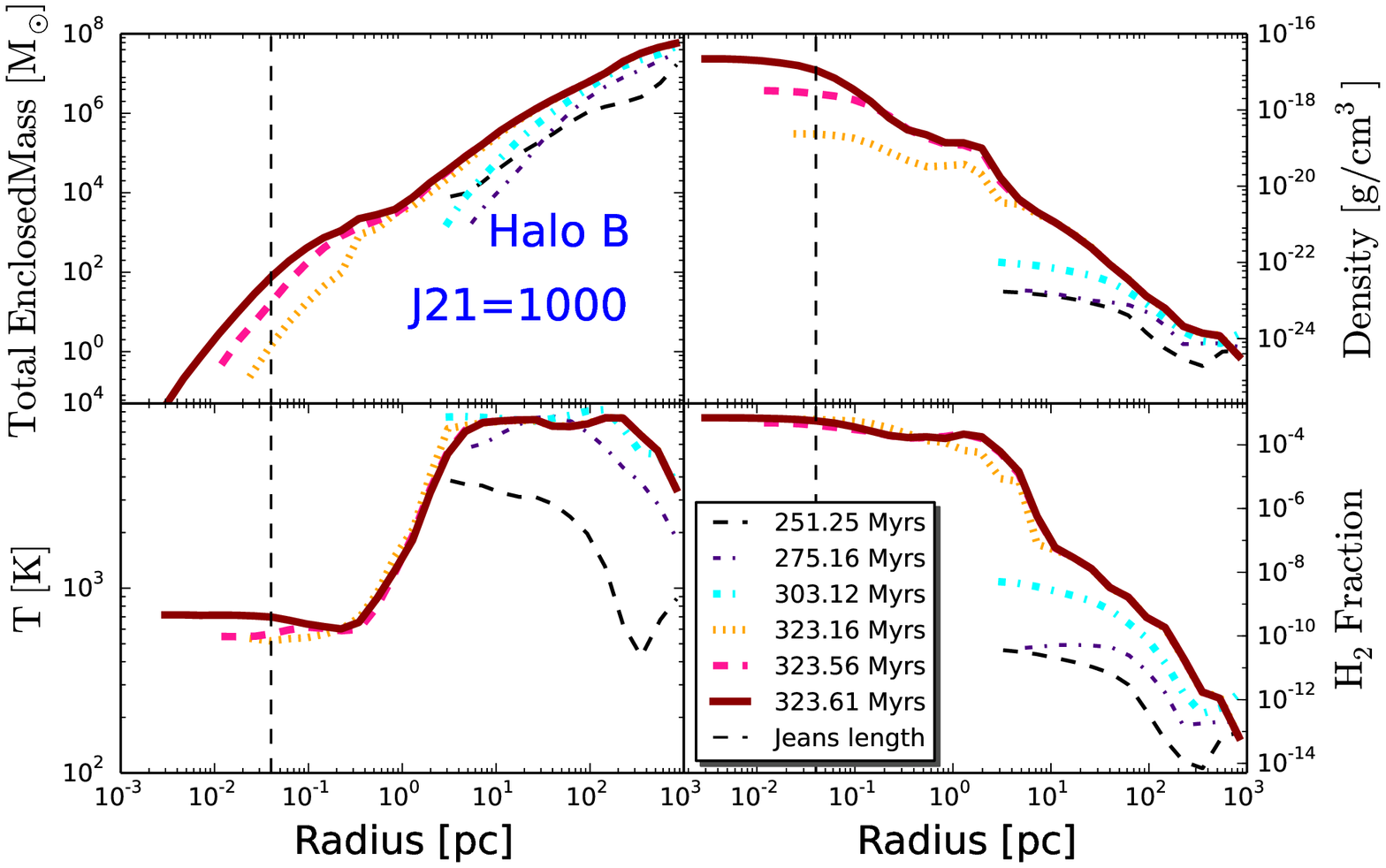} \\
\includegraphics[scale=0.7]{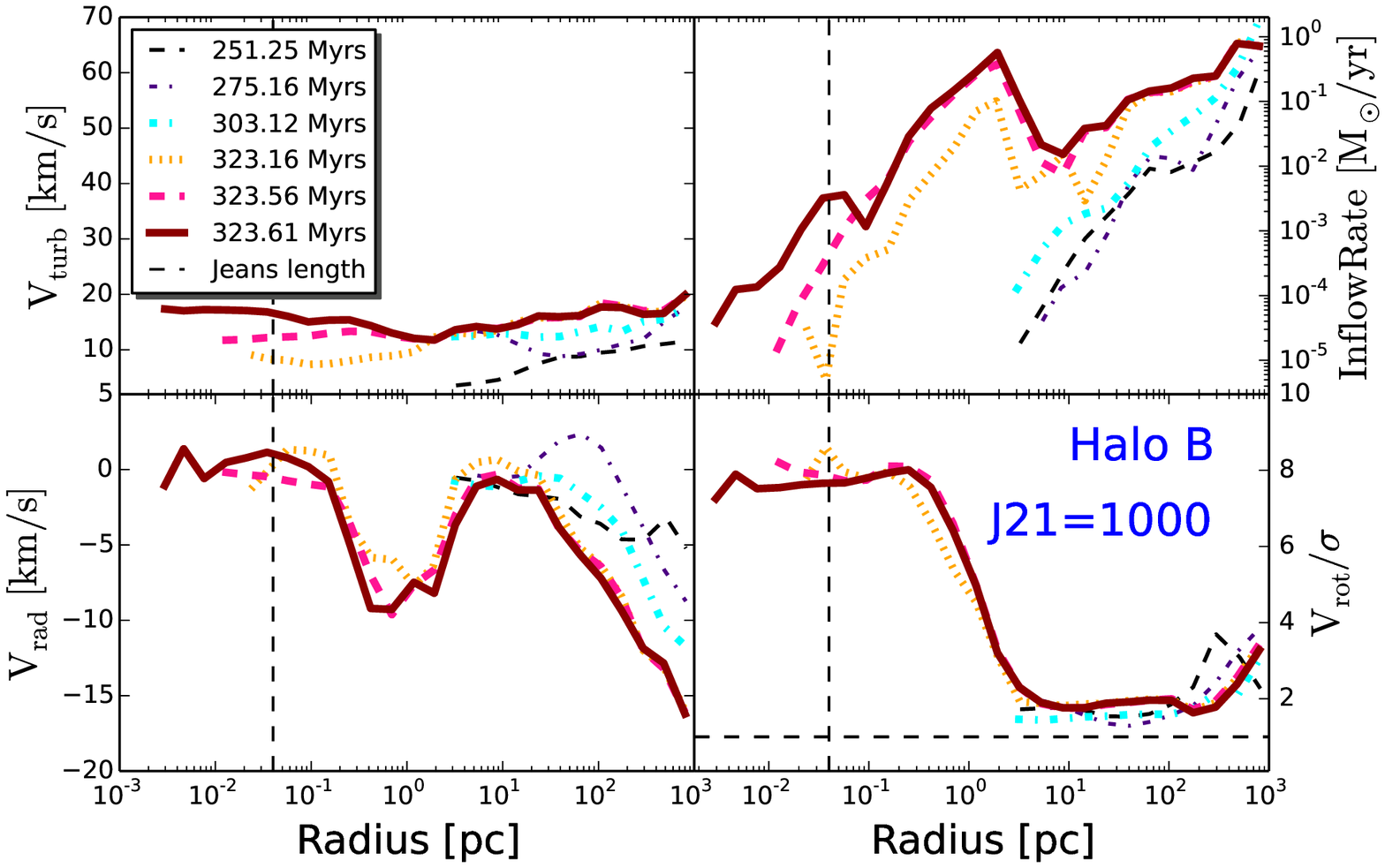} 
\caption{The thermodynamical and physical properties of the halo B are shown for  $\rm J_{21}= 1000$. The top panels show the time evolution of gas density, temperature, $\rm H_2$ fraction and of the total enclosed mass. The  time evolution of the inflow rates, V$_{rot}/ \sigma$, radial infall and turbulent velocities is shown in the bottom panels. The different line styles represent various cosmic times. The profiles are spherically averaged and radially binned. The vertical dashed black line shows the Jeans length and the horizontal dashed line represents V$_{rot}/ \sigma$ =1.}
\label{fig9}
\end{figure*}

\begin{figure*}
\includegraphics[scale=0.7]{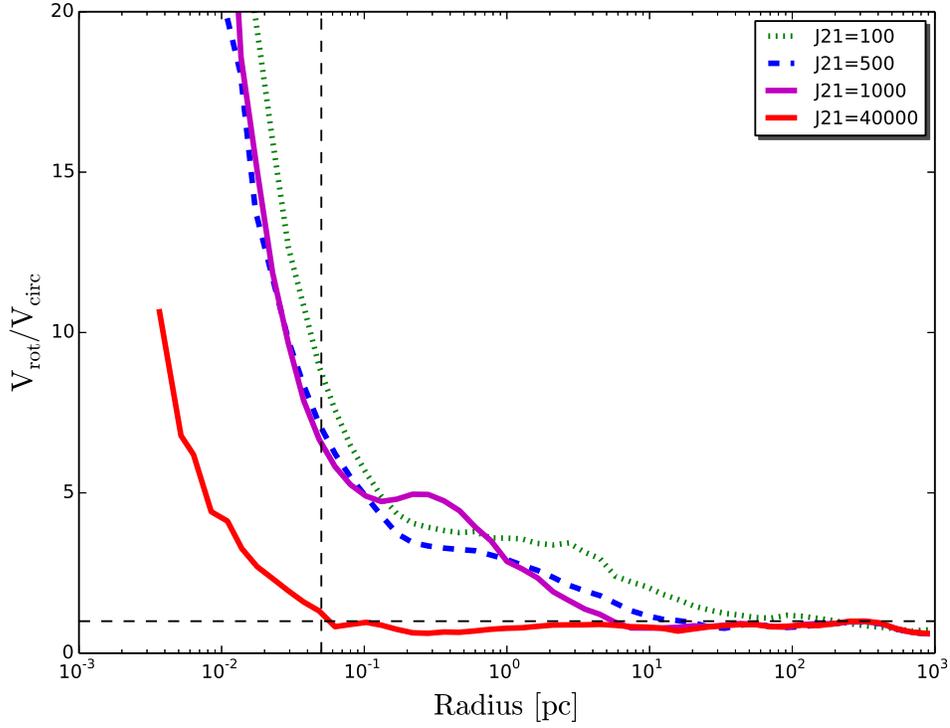} 
\caption{The ratio of rotational to circular velocity is shown here for various BUV fluxes  for halo A, as a representative case.}
\label{figvratio}
\end{figure*}

\begin{figure*}
\includegraphics[scale=0.7]{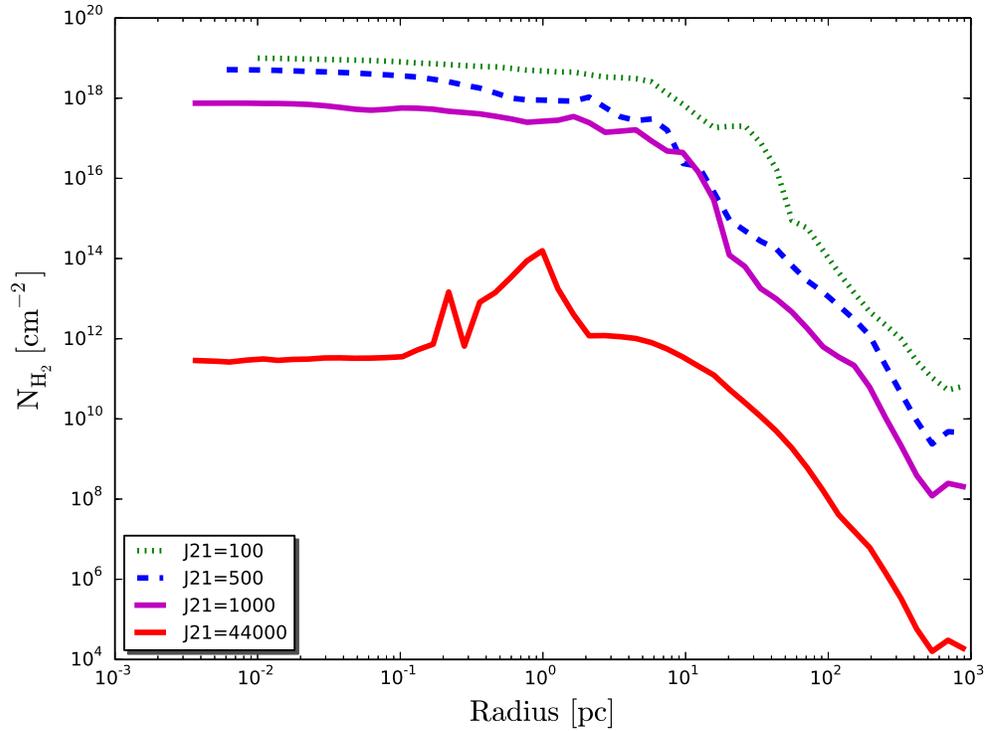} 
\caption{The $\rm H_2$ column density for various BUV fluxes is shown here for  halo A,  as a reference.}
\label{figH2col}
\end{figure*}

For halo B,  the inflow rate for $\rm J_{21}=100$ is between $\rm 0.01 - 0.1~M_{\odot}/yr$ while for  $\rm J_{21}$= 500 and 1000 it is about $\rm 0.1 ~M_{\odot}/yr$ (see figures \ref{fig7}, \ref{fig8}  and \ref{fig9}).  The drop in the inflow rate at  10 pc is due to the formation of an additional clump in the halo. Similar to halo A, the  inflow rate for halo C for $\rm J_{21}$= 100, 500 and 1000 is  $\rm \sim 0.1 ~M_{\odot}/yr$. In all three halos and for all  flux cases, the inflow rate declines within the Jeans length due to the higher thermal pressure. The variations in the inflow rates are correlated with  the radial infall velocity of the halo. The typical infall velocities in these cases are 5-10 km/s. The fluctuations in the infall velocity also indicate the onset of gravitational instabilities which may exert torques and help in the transfer of angular momentum. The infall velocity tends to increase with the strength of the BUV flux and becomes higher particularly for the isothermal collapse. Therefore, the mass inflow rate is larger for the isothermal collapse and is about $\rm 0.1-1~M_{\odot}/yr$. 

\begin{figure*}
\hspace{-9.0cm}
\centering
\begin{tabular}{c}
\begin{minipage}{6cm}
\vspace{0.2cm}
\includegraphics[scale=0.3]{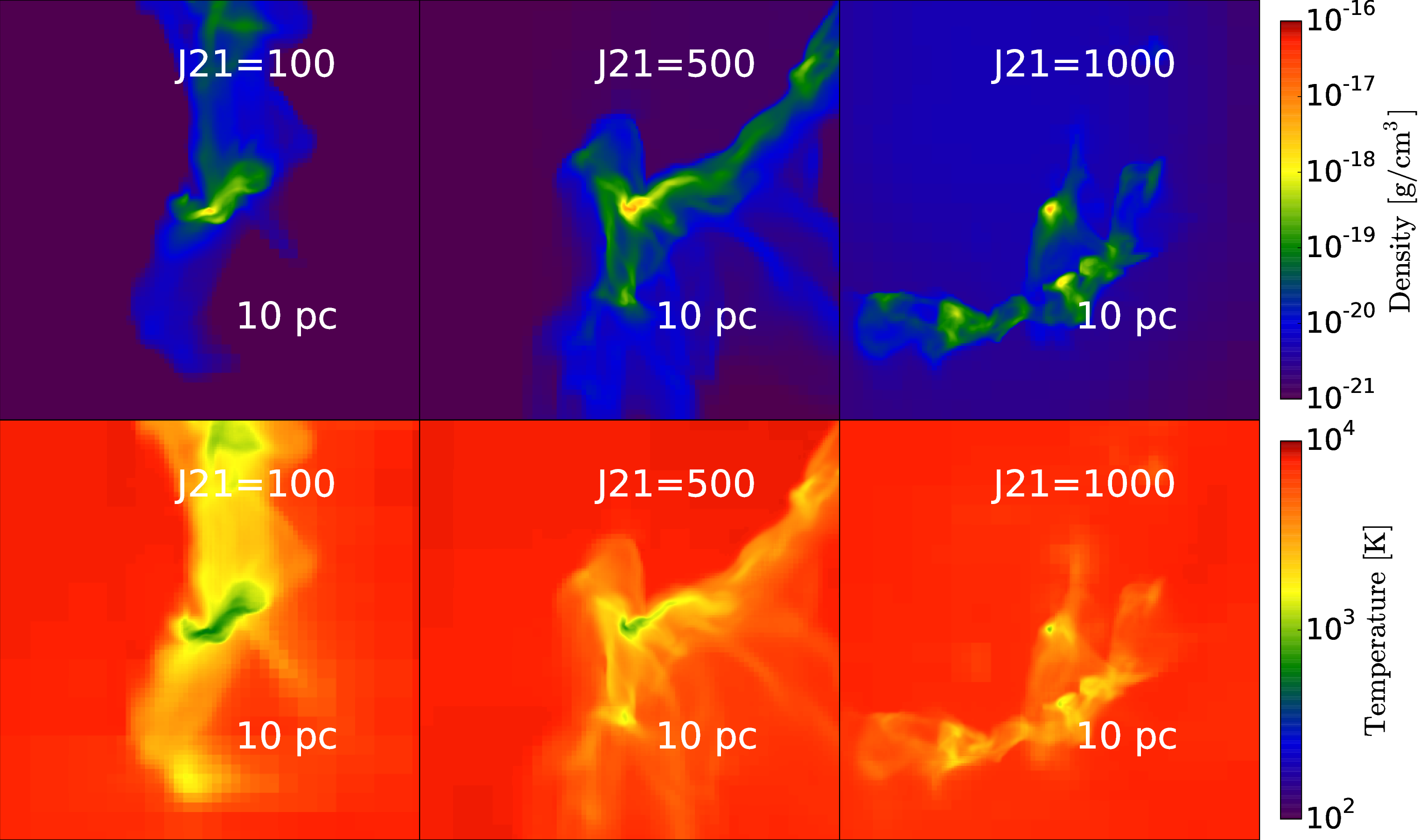} 
\end{minipage} 
\end{tabular}
\caption{ Average density (top panels) and temperature (bottom panels) weighted by the gas density along the y-axis for Halo A.  They are plotted for  the BUV of strengths $\rm J_{21}$ =100, 500 \& 1000 and show the central 10 pc region of the halo when the simulation reaches the maximum refinement level prior to the formation of a sink particle.}
\label{fig19}
\end{figure*}

\begin{figure*}
\hspace{-9.0cm}
\centering
\begin{tabular}{c}
\begin{minipage}{6cm}
\vspace{0.2cm}
\includegraphics[scale=0.3]{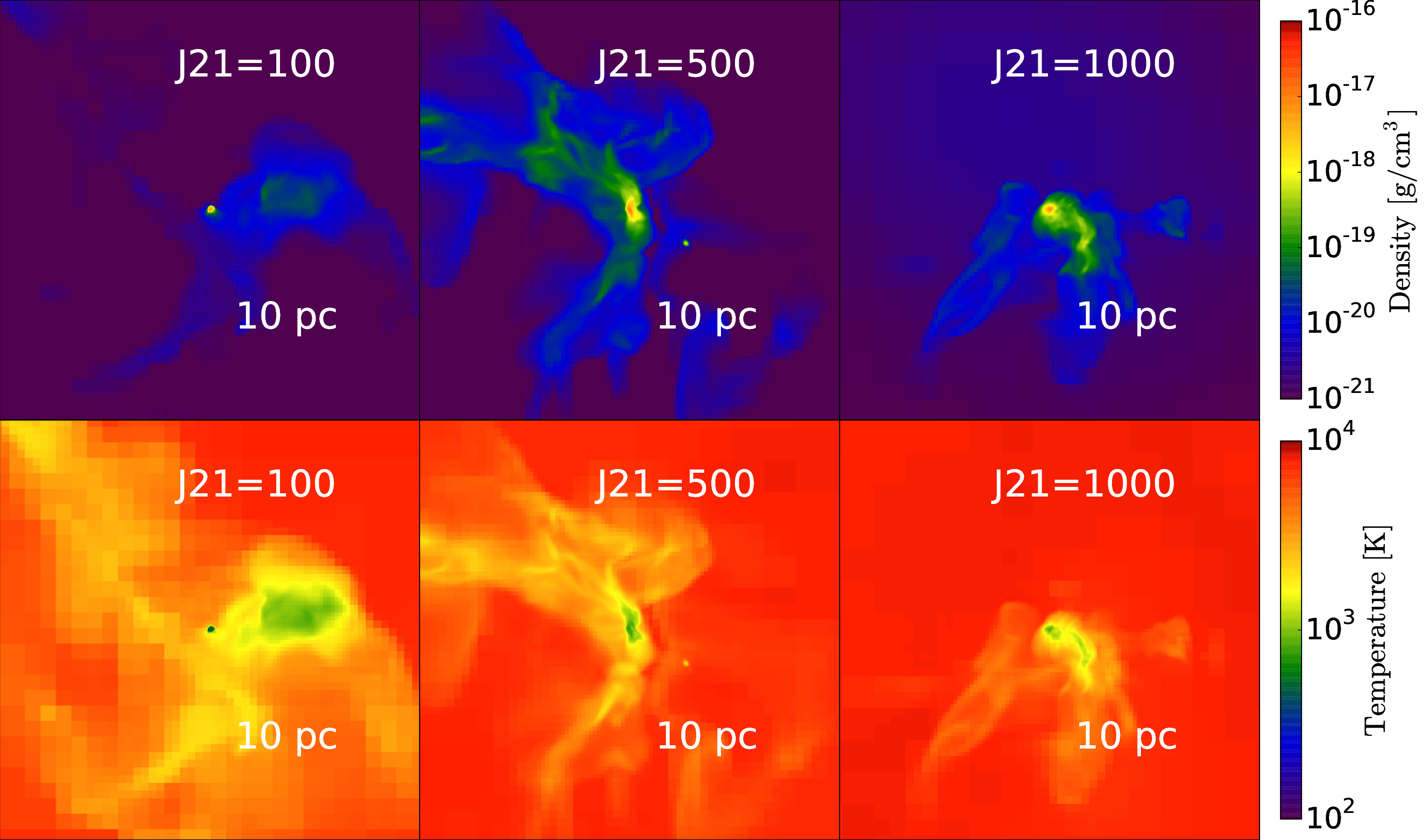} 
\end{minipage} 
\end{tabular}
\caption{Average density (top panels) and temperature (bottom panels) weighted by the gas density along the y-axis for Halo B.  They are plotted for  the BUV of strengths $\rm J_{21}$ =100, 500 \& 1000 and show the central 10 pc region of the halo when the simulation reaches the maximum refinement level prior to the formation of a sink particle.}
\label{fig20}
\end{figure*}

\begin{figure*}
\hspace{-9.0cm}
\centering
\begin{tabular}{c}
\begin{minipage}{6cm}
\vspace{0.2cm}
\includegraphics[scale=0.3]{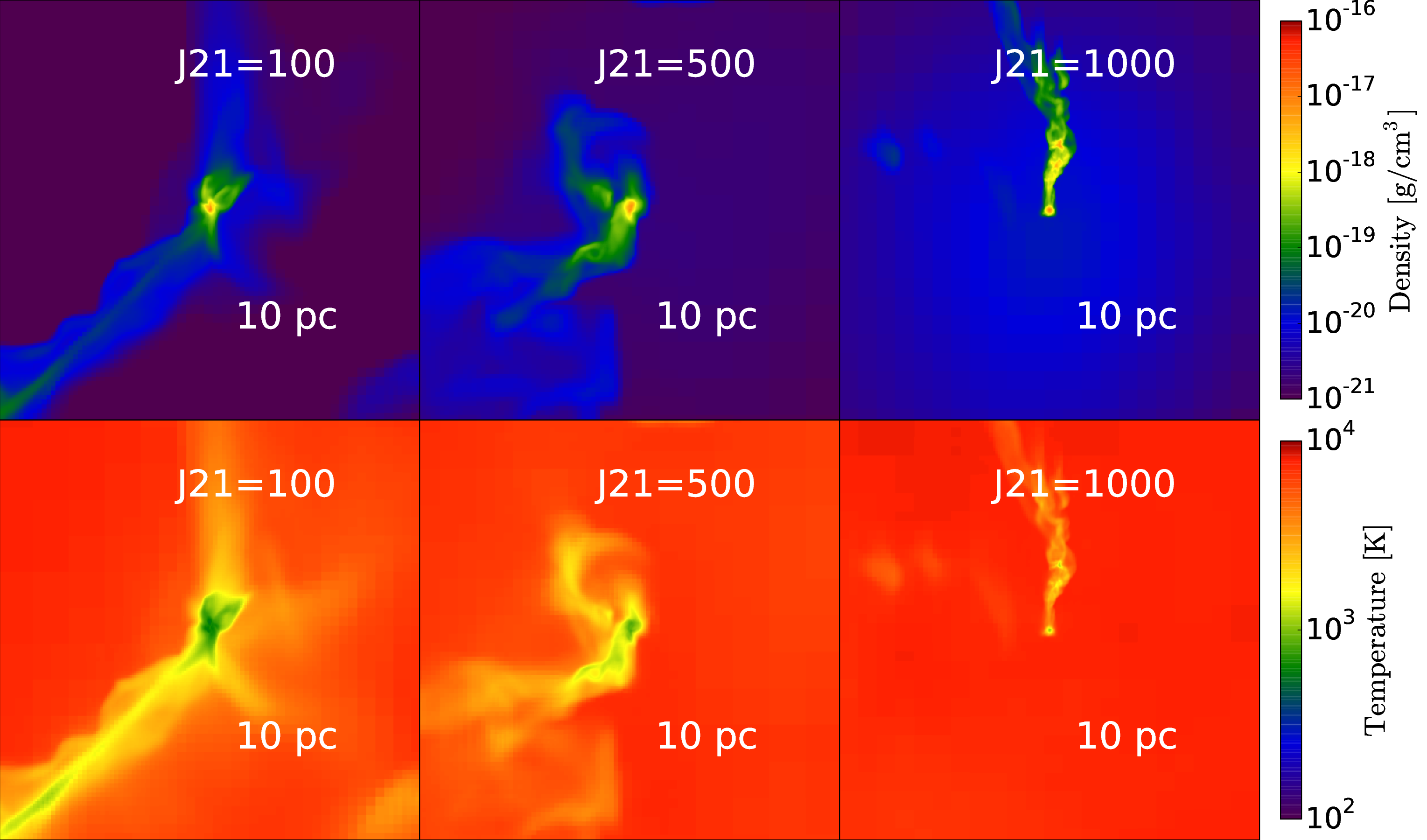} 
\end{minipage} 
\end{tabular}
\caption{Average density (top panels) and temperature (bottom panels) weighted by the gas density along the y-axis for Halo C.  They are plotted for  the BUV of strengths $\rm J_{21}$ =100, 500 \& 1000 and show the central 10 pc region of the halo when the simulation reaches the maximum refinement level prior to the formation of a sink particle.}
\label{fig21}
\end{figure*}

We also estimated the total (DM + gas) enclosed mass in  halo A. The enclosed mass within the central 10 pc of the halo is about a factor of a few higher for  $\rm J_{21}=1000$ compared to the $\rm J_{21}=100$ as shown in figures \ref{fig7}, \ref{fig8}  and \ref{fig9}. The mass profiles show a plateau around a few thousand solar masses which indicate the formation of small scale disk, comparable to the  estimates of $V_{rot}/ \sigma$ and also found by \cite{Regan2014B}. This plateau becomes even more visible for halo B in the $\rm J_{21}=100$ case. We also estimated the turbulent velocity for halo A  in the presence of various strengths of UV flux and is shown in  figures \ref{fig3}, \ref{fig4}, \ref{fig5}  and  \ref{fig6}. The typical turbulent velocity  is about 10 km/s and fluctuations in the turbulent velocity at various radii are related to the density structure and intermittent flows occurring. For the other two halos, we found similar trends both for the enclosed mass and turbulent velocities. Again,  small variations in these quantities are observed from halo to halo. The gaseous structures (density and temperature) within the central 10~pc of the three halos are shown in figures~\ref{fig19}, \ref{fig20}, \ref{fig21}. Either extended (halo A and halo B) or narrow funnels of dense gas connect to the halo centre, providing the high inflow rates we measured.

\subsection {Estimates of the masses of supermassive stars}
We selected two out of three halos, those with the highest inflow rates, to study the long term evolution of mass accretion rates. We here only focus on the weaker BUV flux cases, where collapse is not fully isothermal. It has been shown that strong BUV fluxes lead to complete isothermal collapse where large accretion rates of about $\rm 1~M_{\odot}/yr$ can be obtained and massive objects of $\rm 10^5~M_{\odot}$ can be formed \citep{Regan09,Latif2013d,Bcerra2014}. However, as explained earlier, these sites are very rare.  As mentioned in the previous subsection, in halos A and C mass inflow rates of $\rm 0.1~M_{\odot}/yr$ can be obtained for  $\rm J_{21} \geq 500$ and even for $\rm J_{21} =100$.  To  assess if these  rates can be maintained for long time scales, we employed sink particles in our simulations and followed accretion onto them for about 30000 years after the formation of the first sink particle. The sinks are formed at densities above $\rm 10^7~cm^{-3}$ and at the maximum refinement level. In general, no vigorous fragmentation is observed in either halo for all strengths of BUV flux. Only a single massive sink forms per halo.  

\begin{figure*}
\hspace{-9.0cm}
\centering
\begin{tabular}{c}
\begin{minipage}{6cm}
\vspace{0.2cm}
\includegraphics[scale=0.3]{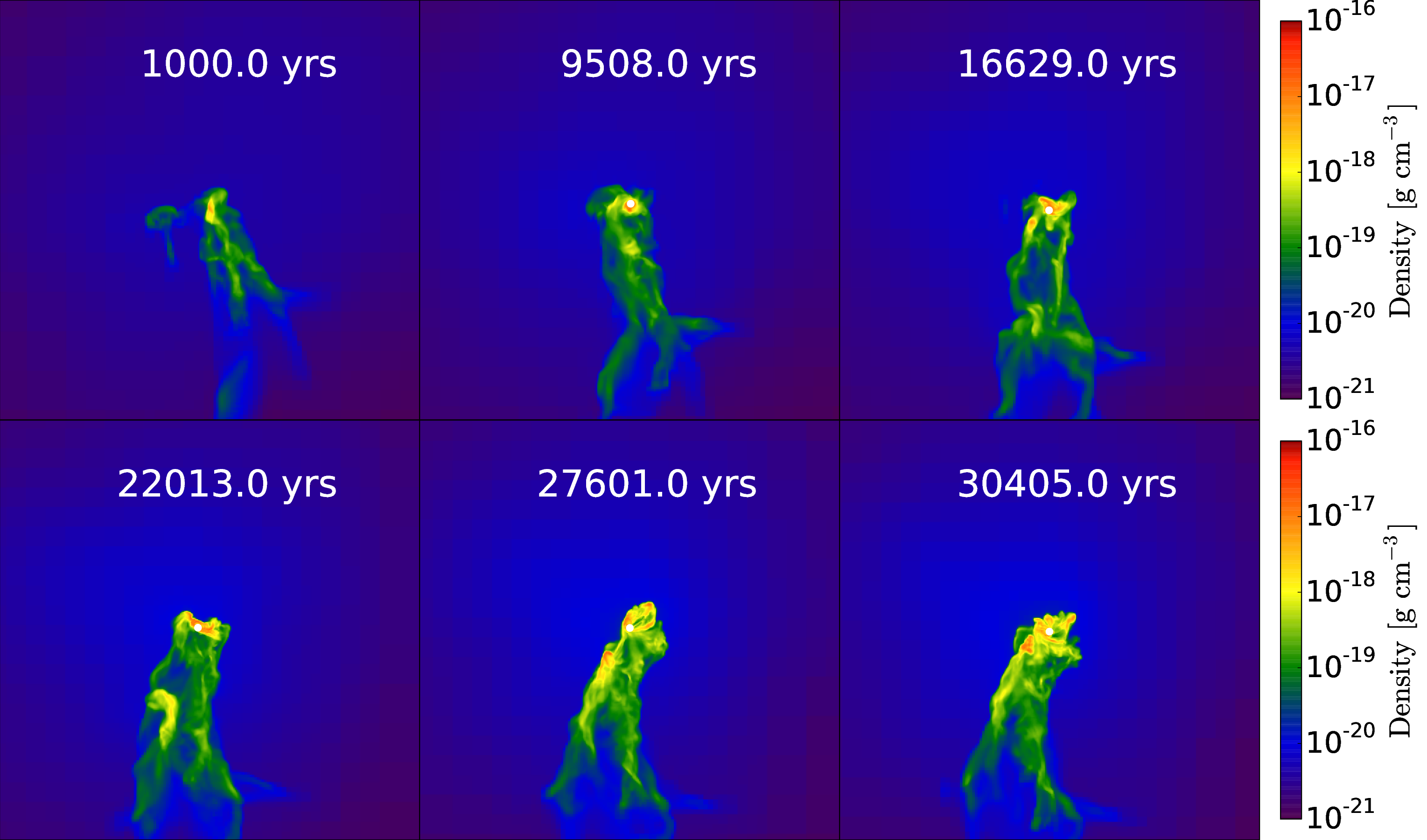} 
\end{minipage} 
\end{tabular}
\caption{The time evolution of the density averaged along the y-axis is shown in the central 10 pc of the halo A for $\rm J_{21}$=1000 after the formation of the sink particle.  The  time after the formation of the sink particle in each panel is shown in years and the white dot depicts the position of the sink particle.}
\label{fig22}
\end{figure*}

\begin{figure*}
\hspace{-9.0cm}
\centering
\begin{tabular}{c}
\begin{minipage}{6cm}
\vspace{0.2cm}
\includegraphics[scale=0.3]{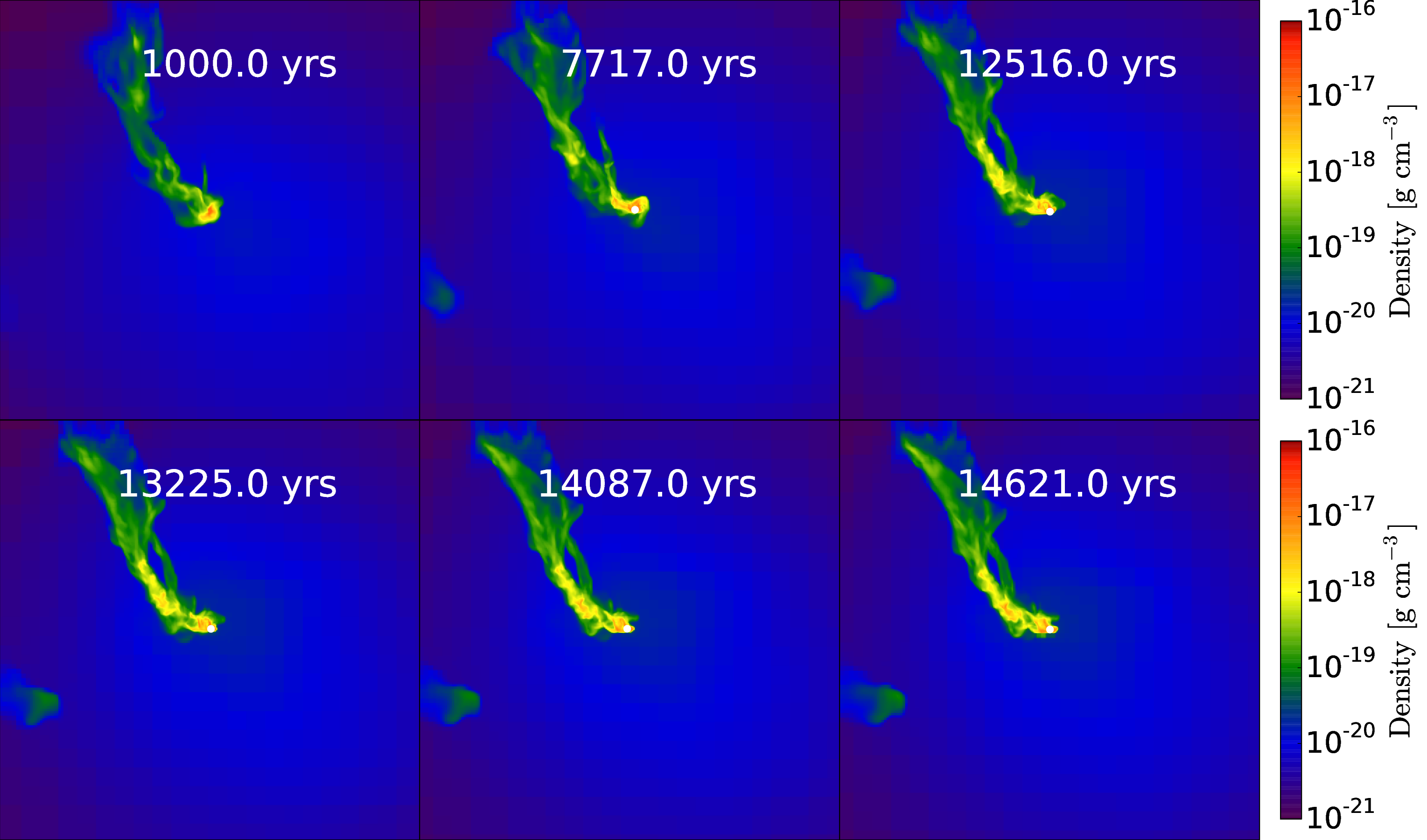} 
\end{minipage} 
\end{tabular}
\caption{The time evolution of the density averages along the y-axis is shown in the central 10 pc of the halo C for $\rm J_{21}$=1000 after the formation of the sink particle.  The  time after the formation of the sink particle in each panel is shown in years and the white dot depicts the position of the sink particle.}
\label{fig29}
\end{figure*}

Figures~\ref{fig22} and~\ref{fig29} show the time evolution of the gas density  in the central 10 pc of halos A and C for $\rm J_{21}=1000$ after the formation of the sink particle. Note the highly anisotropic distribution of gas: the central sink is mainly fed by a narrow stream of gas, rather than by an extended, spherical or axisymmetric structure as in the case of a higher UV flux, where collapse is isothermal \citep[cf. figures 2 and 8 in][]{Latif2013d}. 

\begin{figure}
\includegraphics[width=\columnwidth]{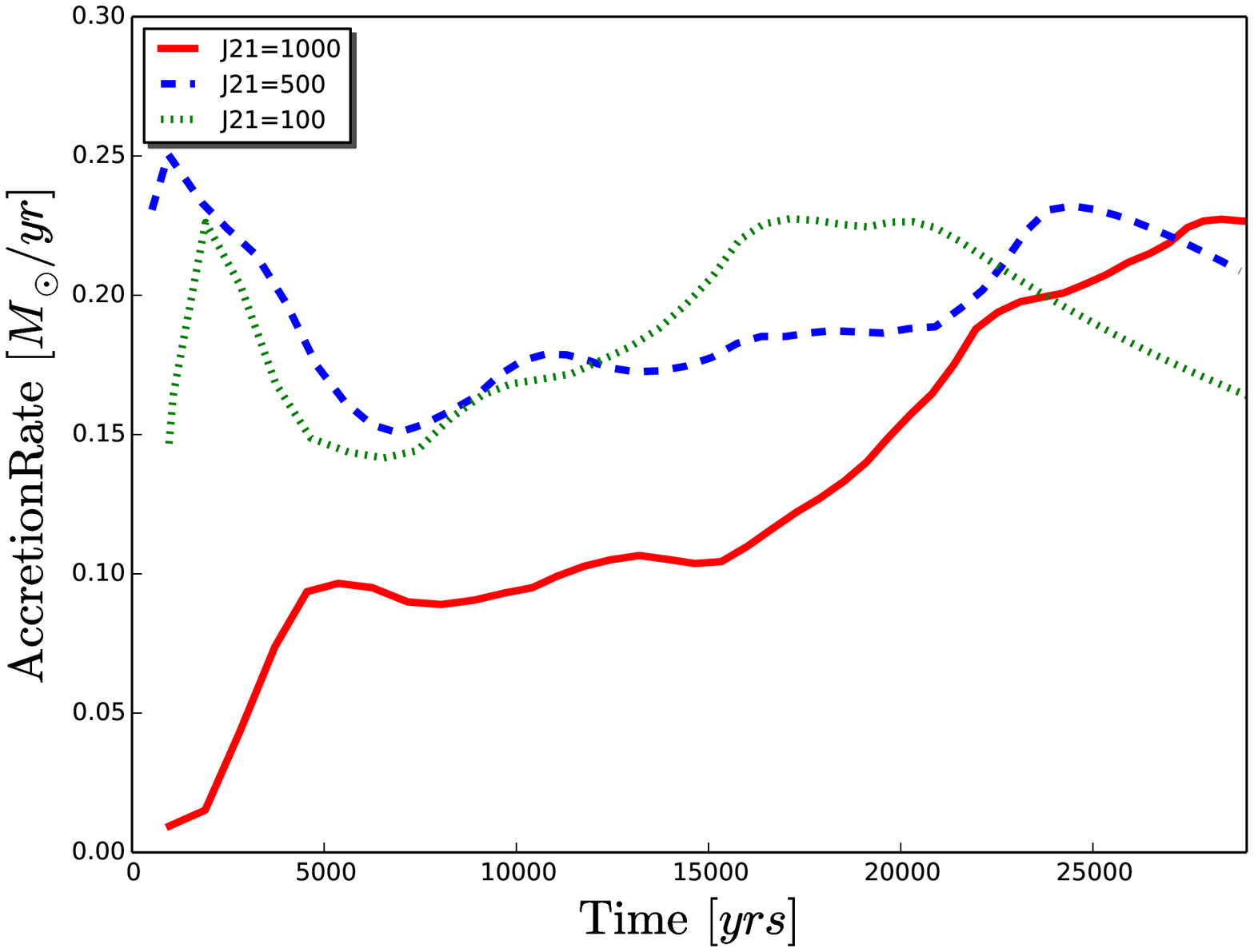} \hfill
\includegraphics[width=\columnwidth]{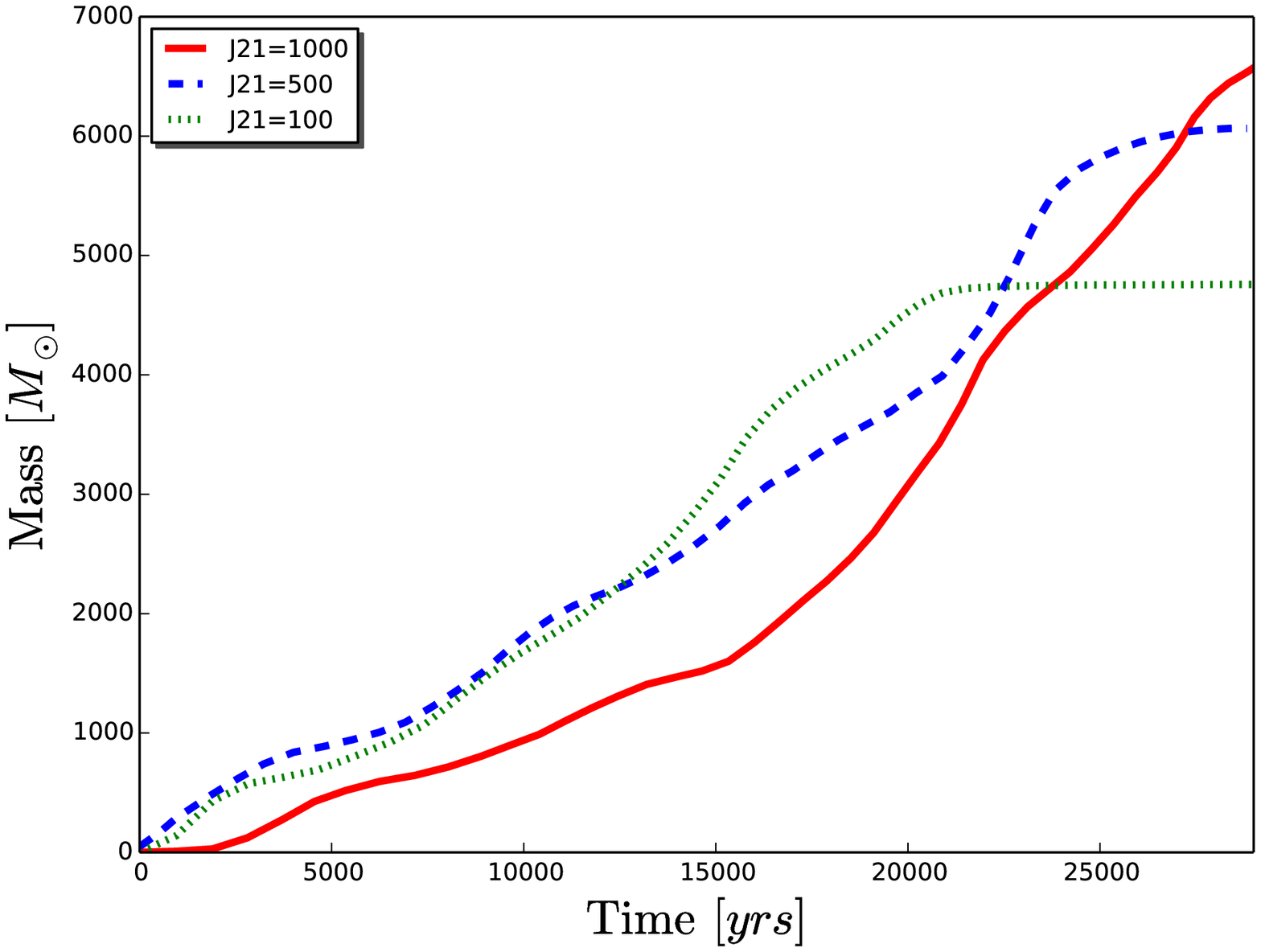} 
\caption{The masses of the sink particles and their mass accretion rate are shown against time for  halo A. The strength of incident field is shown by  different colors and line styles as described in the legend.}
\label{fig16}
\end{figure}

The temporal evolution of  the mass accretion rates onto the sinks for halo A, illuminated by  various strengths of BUV flux,  is shown in figure \ref{fig16}.  A mass accretion rate $\rm \geq 0.1~ M_{\odot}/yr$ can be maintained for the simulated time for all cases.  In the $\rm J_{21}=1000$ case,  the gas distribution  at the time the sink particle is inserted is less dense and more extended,  and it  takes about 10000 yrs to  move the smaller clumps into the centre and, after that, the inflow rate increases. Additionally, the small clumps that form,  eventually merge in the centre. Therefore, in the last 5000 yrs of the simulated time, the mass accretion rate increases further for $\rm J_{21}=1000$ while the mass accretion rate for  $\rm J_{21}=100$ and 500 cases declines.

The masses of the sinks after  30000 years, are $\rm 4700, ~6000 ~ and ~ 6770~M_{\odot}$ for  $\rm J_{21}=100$, 500 and 1000, respectively. The mass of the sink is 2000 solar masses higher for $\rm J_{21}=1000$ compared to the weakest flux case. In fact, the mass for $\rm J_{21}=100$ shows a plateau around 4700 solar masses and  becomes almost constant for the last 10000 years.  This is because of the higher rotational support as well as  the presence of colder gas compared to the higher flux cases. The small plateau in the sink mass and  a small decline in the mass accretion rate are also observed for $\rm J_{21}=500$ in the last 5000 yrs. However, the mass of the sink seems to keep increasing for the $\rm J_{21}=1000$ case. Therefore, we expect that the masses of the resultant star may be higher compared to the lower flux cases.

\begin{figure}
\includegraphics[width=\columnwidth]{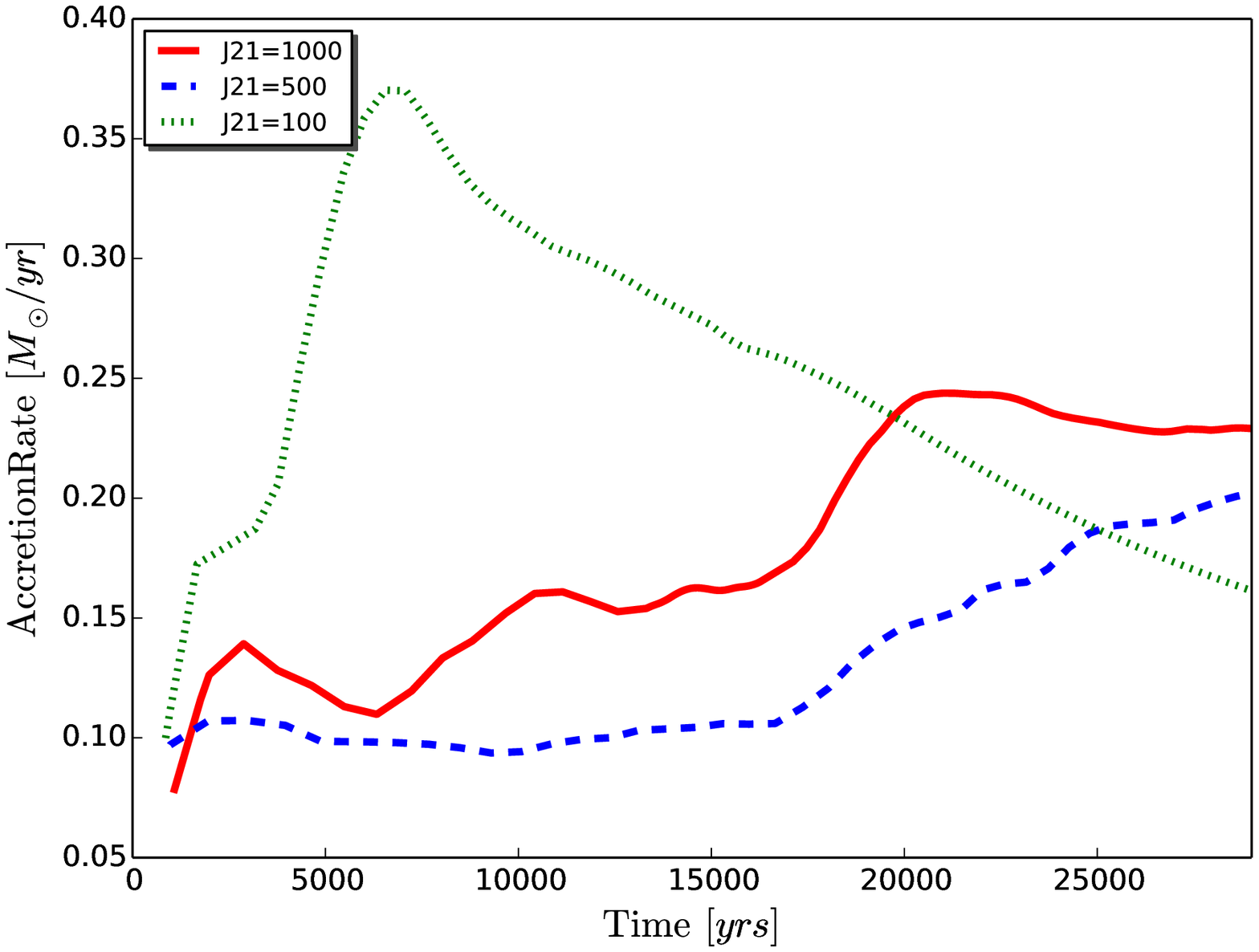} \hfill
\includegraphics[width=\columnwidth]{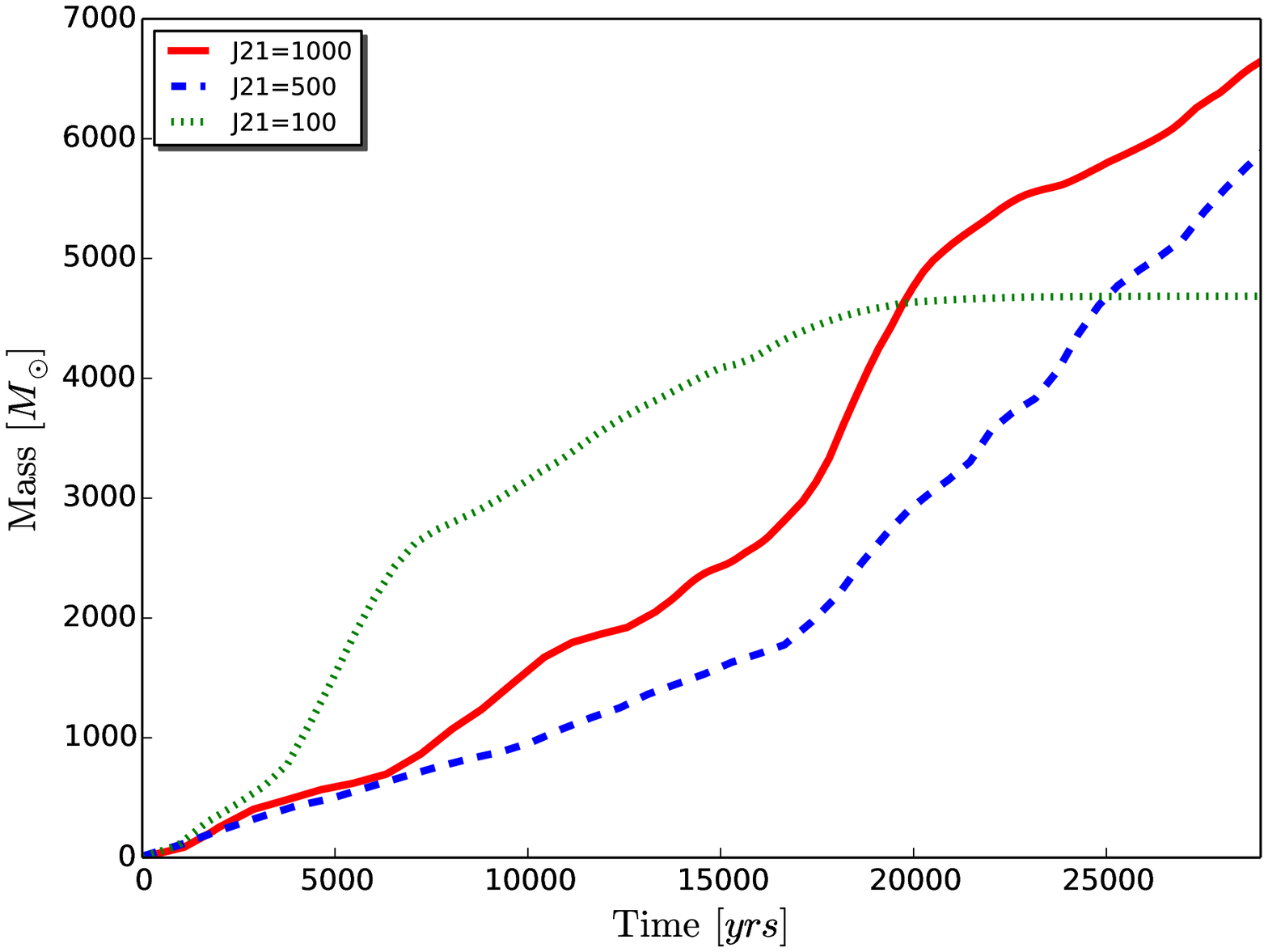} 
\caption{The masses of the sink particles and their mass accretion rate are shown against time for  halo C. The strength of incident field is shown by  different colors and line styles as described in the legend. }
\label{fig17}
\end{figure}

We have also employed sink particles for halo C and followed the evolution for about 30000 yrs. The masses of the sinks and the mass accretion rates onto them are shown in figure \ref{fig17}. The mass accretion rate starts to increase in the first 10000 yrs due to the sufficient mass supply in the central clump, but decline later. This behaviour is similar to halo A, but the decline is more significant. The sink mass starts to show a plateau  earlier, at about 18000 yrs after the formation of the sink. The mass accretion rates and the masses of the sink particles continue to increase for the rest of the two cases. The masses of the sinks are 4600, 6400 and 7000 $\rm M_{\odot}$  from the weaker to stronger fluxes,  respectively (i.e. $\rm J_{21}$ = 100, 500 and 1000).  Similar to halo A,  mass accretion rates remain above $\rm 0.1~M_{\odot}/yr$ for the simulated time, irrespective of the BUV field strength.  The mass accretion rates show some decline for $J_{21}$=100 case, but not for the $\rm J_{21}$= 500 and 1000 cases.  Mass accretion rates may decline at  later stages of evolution if rotational support is enhanced significantly, or efficient fragmentation takes place inside the halo.   However, for the duration of our simulation, we confirm $\rm \geq 0.1~ M_{\odot}/yr$  for the whole time. 

Overall, our simulations suggest that the mass accretion rates required to build a supermassive star can be obtained for  UV fluxes below the critical value  necessary for isothermal collapse. Particularly, the formation of supermassive stars seems plausible for  $\rm J_{21}=1000$ and 500, at least in one case. If accretion persists at the same rates, the masses of the sinks can reach $\rm 10^4-10^5$ solar masses.

\section{Discussion AND conclusions}

We have performed high resolution cosmological simulations for three distinct halos to study under which conditions the accretion rates necessary for  the formation of supermassive stars in  massive primordial halos of a few times $\rm 10^7 ~M_{\odot}$ can be achieved. These supermassive stars are the potential cradles of direct collapse black holes which are  one of the  candidates to explain the existence of high redshift quasars. 

The prime objective of this work is to explore the formation of supermassive stars in non-isothermal collapse. Indeed, the key parameter for supermassive star formation appears to be an inflow rate above $\rm 0.1 ~M_{\odot}/yr$ \citep{Begelman2010,Hosokawa2013,Schleicher13,Ferrara14}. Specifically, we are interested in estimating whether the concept of a critical value, $J_{crit}$, of the BUV flux above which full isothermal collapse can be achieved through complete suppression of molecular hydrogen formation is necessary for black hole formation in the direct collapse scenario. 

We also study the structure of the inner regions of the halos to evaluate the role of rotational support. We presume that halos are metal free and illuminated by moderate strengths of BUV flux.  A pc-size fat disk is assembled in these halos  as a consequence of gravitational collapse and the angular momentum of the infalling gas.  The formation of such disk is irrespective of the merger history of the halo. The scale height of the disk  varies from  $\rm 0.25 - 1$ depending on the thermodynamical properties of the halo.  Halos cooled by the molecular hydrogen  have higher rotational support and an aspect ratio lower than  the halos collapsing isothermally  \citep{Oh2002,Lodato2006,Begelman2006}. 

Rotational support, however, does not, at least initially, hinder gas collapse in any of the simulated halos. In fact, our estimates for the mass inflow rates show that the halos irradiated by moderate UV fluxes,  $\rm J_{21}$= 500 and 1000 have  typical  mass inflow rates of about $\rm 0.1 ~M_{\odot}/yr$.  Indeed, \cite{Regan2014B} already noticed that sufficiently high accretion rates can be obtained for halos experiencing (anisotropic) fluxes of $\rm J_{21}\sim 1000$.  An inflow rate of  $\rm \sim 0.1 ~ M_{\odot}/yr$ can be achieved even for $\rm J_{21} =100$  for halos  A and C, while halo B has somewhat lower inflow rates.  This may  suggest that halos experiencing major mergers in their recent past may form supermassive stars more easily,  for lower strengths of UV field, while perhaps this is not happening in  haloes formed via minor mergers and accretion.  In fact,  \cite{Mayer2014} suggested  that the merging of very massive galactic cores, $>\rm 10^9~ M_{\odot}$, may form a stable nuclear disk with very large gas inflow rates. They argue that merging of such systems leads to the enhanced inflow rates and helps in the formation of a massive black hole.  Although the present work explores very different systems,  the merging of a few times $\rm 10^6~ M_{\odot}$ proto-galaxies, it hints that  larger accretion rates seem to be obtained  in the aftermath of a major merger, as mergers induce torques that remove angular momentum from gas, favoring inflows. However, our small sample does not allow us to draw strong conclusions.  \cite{Inayoshi2015} have proposed an alternative scenario which does not require the  BUV to photo-dissociate molecular hydrogen but the collisions of protogalaxies with velocities of a few hundred km/s  produce strong shocks with T $\rm > 10^6$ K where subsequently gas collapses isobarically and $\rm H_2$ gets collisionally dissociated.

To further assess whether such higher accretion rates can be maintained for  long time scales, we have employed sink particles and followed accretion on them for 30000 years after the formation of the first sink particle for halos A and C. No strong fragmentation is found for these halos, irrespective of the employed UV field, and a single star is formed per halo. A mass inflow $\rm \sim 0.1 ~ M_{\odot}/yr$ can be maintained for long times, at least for the whole simulated time.  We stopped the simulations  30000 years after the formation of the first sink as evolving them further becomes computationally very expensive. However, given these accretion rates, supermassive stars may reach $\rm 10^4 - 10^5~M_{\odot}$ within about one million years. Our results suggest that $\rm J_{21} \geq  500 - 1000$ may be sufficient to form supermassive stars, at least for some cases.  The observed abundance of $z  \geq 6$ quasars can be reproduced by the above flux,  extrapolating the findings of \cite{Dijksta2014}. 

In our simulations, we employed sink particles to follow the mass accretion onto them, which introduce a characteristic length scale below which fragmentation is ignored.  To assess  fragmentation below this scale would require simulations resolving the collapse down to sub AU scales and it becomes computationally impractical  to evolve them for longer times. This is due to the shorter time-steps involved in these calculations which cover a  large range of spatial scales, from cosmological scales down to the scales of AU. However, the analysis of the density structure in these halos indicates that no vigorous fragmentation is expected, as $\rm H_2$ cooling is only effective in the core of the halo and most of the gas outside the core is warm and no signs of substructures are present.  Even if fragmentation occurs on very small scales, \cite{LatifViscous2014,Latif2015Disk} show that clumps may migrate inward on short time scales. Moreover, viscous heating further dissociates molecular hydrogen, stabilises the disk and favours the formation of a massive central object. Based on the results from both these simulations and the complementary analytical studies \citep{LatifViscous2014,Latif2015Disk}, we argue that formation of supermassive stars in these conditions is possible.

We also found that accretion onto the central star becomes highly anisotropic particularly for the case with $\rm J_{21}=1000$ where the gas is fed to the central star by a single stream. Such behaviour seems to persist for 30000 yrs. This may have important implications for feedback by the central star or even by the black hole at later stages. The ionising radiation around a star/black hole would escape in the low density region while the gas supply to the central object may continue from a single dense stream even in the presence of  feedback. 

The simulations presented in this work are based on a number of idealized assumptions such as halos are metal  free and irradiated by an isotropic flux. In the  future, simulations self-consistently taking into account  metal enrichment and radiation emitted by the nearby sources should be performed.  Such simulations are mainly constrained by the exorbitant computational costs.

 \newpage
\clearpage

\newpage
\clearpage

 \section*{Acknowledgments}
The research leading to these results has received funding from the European Research Council under the European Community's Seventh Framework Programme (FP7/2007-2013 Grant Agreement no. 614199, project ``BLACK'').  This work was granted access to the HPC resources of TGCC under the allocation x2015046955 made by GENCI. The simulation results are analyzed using the visualization toolkit for astrophysical data YT  \citep{Turk2011}.

 \bibliography{smbhs.bib}
 
\newpage

\end{document}